\documentclass[12pt, /preprint]{aastex}
\def\be{\begin{equation}}
\def\ee{\end{equation}}

\def\etal{{\it et al. }}
\def\hi{{\ion{H}{1}}}

\newcommand{\kms}{\ensuremath{\rm{km\,s^{-1}}}}

\newcommand{\dg}{\ensuremath{^{\circ}}}
\newcommand{\jykms}{\ensuremath{\mathrm{Jy} \,\kms}}
\newcommand{\cm}{\ensuremath{\mathrm{cm}^{-2}}}
\def\msun{\ensuremath{M_\odot}}

\def\dv{\ensuremath{D_V}}
\def\sn{\ensuremath{S/N}}
\def\nh{\ensuremath{N_{HI}}}
\def\nhp{\ensuremath{\nh^{\prime}}}
\def\chan{\ensuremath{\delta V}}
\def\chanp{\ensuremath{\chan^{\prime}}}
\def\sm{\ensuremath{\sigma_{m}}}
\def\smp{\ensuremath{\sm^{\prime}}}
\def\mhi{\ensuremath{M_{HI}}}
\def\mhip{\ensuremath{\mhi^{\prime}}}
\def\wf{\ensuremath{W50}}
\def\wfp{\ensuremath{\wf^{\prime}}}
\def\fc{\ensuremath{F_c}}
\def\fcp{\ensuremath{\fc^{\prime}}}
\def\cz{\ensuremath{cz}}
\def\czsun{\ensuremath{\cz_{\odot}}}
\def\czsunp{\ensuremath{\czsun^{\prime}}}
\def\mdyn{\ensuremath{M_{dyn}}}
\def\mdynp{\ensuremath{\mdyn^{\prime}}}
\def\ahi{\ensuremath{a_{HI}}}
\def\ahip{\ensuremath{\ahi^{\prime}}}
\def\pahip{\ensuremath{PA_{HI}^{\prime}}}

\def\apjl{{ApJL}}
\def\apj{{ApJ}}
\def\apjs{{ApJS}}
\def\aj{{AJ}}
\def\aap{{A\&A}}

\def\mnras{{MNRAS}}

\usepackage{epsfig}

\begin{document}



\title{The Arecibo Legacy Fast ALFA Survey VII:  A Neutral Hydrogen Cloud Complex in the Virgo Cluster}
\author {Brian R. Kent\altaffilmark{1,2}, Kristine Spekkens\altaffilmark{3}, Riccardo Giovanelli\altaffilmark{4,5}, Martha P. Haynes\altaffilmark{4,5}, 
Emmanuel Momjian\altaffilmark{6}, Juan R. Cort{\'e}s\altaffilmark{7,8}, Eduardo Hardy\altaffilmark{9,7}, and Andrew A. West\altaffilmark{10}
}

\altaffiltext{1}{Jansky Fellow of the National Radio Astronomy Observatory.}

\altaffiltext{2}{National Radio Astronomy Observatory, 520 Edgemont Road, Charlottesville, VA 22903.
The National Radio Astronomy Observatory is a facility of the  
National Science Foundation operated under cooperative agreement by  
Associated Universities, Inc.
{\it e--mail:} bkent@nrao.edu}

\altaffiltext{3}{Department of Physics, Royal Military College of Canada, P.O. Box 17000, Station Forces,
Kingston, Ontario, K7K 7B4. {\it e--mail:} Kristine.Spekkens@rmc.ca}

\altaffiltext{4}{Center for Radiophysics and Space Research, Space Sciences Building,
Cornell University, Ithaca, NY 14853. {\it e--mail:} riccardo@astro.cornell.edu,
haynes@astro.cornell.edu}

\altaffiltext{5}{National Astronomy and Ionosphere Center, Cornell University,
Space Sciences Building,
Ithaca, NY 14853. The National Astronomy and Ionosphere Center is operated
by Cornell University under a cooperative agreement with the National Science
Foundation.}

\altaffiltext{6}{National Radio Astronomy Observatory,
1003 Lopezville Rd.,
P. O. Box O,
Socorro, NM 87801.
{\it e--mail:} emomjian@nrao.edu
}

\altaffiltext{7}{Departamento de Astronom{\'i}a, 
Universidad de Chile, 
Casilla 36-D, 
Santiago, Chile
{\it e--mail:}  jcortes@das.uchile.cl
}

\altaffiltext{8}{National Astronomical Observatory of Japan,
2-21-1 Osawa, Mitaka, Tokyo, 181-8588
}

\altaffiltext{9}{National Radio Astronomy Observatory, 
Casilla El Golf 16-10, 
Las Condes, Santiago, Chile
{\it e--mail:} ehardy@nrao.edu
}

\altaffiltext{10}{Astronomy Department,
601 Campbell Hall,
University of California,
Berkeley, CA 94720-3411
{\it e--mail:} awest@astro.berkeley.edu
}

\begin{abstract}
We present observations of an HI cloud complex most likely located in the Virgo galaxy 
cluster, first reported by Kent \etal (2007). The complex consists of five clouds, 
detected in the data set of the ALFALFA extragalactic HI survey at Arecibo. The clouds 
have radial velocities between $cz_{\odot}~\sim$480 and 610 \kms. At the Virgo cluster
distance, they are spread over a projected span of 170 kpc and have HI masses ranging  
from 0.48 to 1.7 $\times 10^8$ \msun. The overall HI mass of the complex is 
$5.1\times 10^8$ \msun. The clouds' velocity 
widths vary between 50 and 250 \kms. New results of follow-up aperture synthesis 
observations conducted with the Very Large Array are also presented, which yield
a higher resolution view of two of the clouds in the complex. These two resolved
clouds show no evidence of symmetry in the gas distribution or of any ordered
motions. The possibility that the complex is a group of primordial objects, embedded
in their own dark matter halo is thought to be unlikely. Scenarios in which the
clouds have been removed from the disk of a galaxy traveling at high speed through
the intracluster medium are considered. The most likely among those is thought to be
one where the clouds were separated from NGC~4445 at a time $>0.5$ Gyr ago. The 
orbital velocity of the clouds and the putative parent galaxy would now be seen at 
a relatively large angle with respect to the line of sight.
\end{abstract}


\section{Introduction}
\label{intro}

The study of the cold gas content in extragalactic systems plays an important role in 
understanding their formation and evolution.  
Not only does the gas content reveal information
on the star formation potential of a galaxy, it can also help trace
the dynamic history and response to the surrounding
environment, whether it inhabits a dense cluster or an unpopulated field.
For example, the fingerprint of galactic collisions and ram pressure stripping effects can 
be seen in the form of detached gaseous fragments 
and clumps. The 21 cm spectral line of neutral hydrogen 
serves as an important tool for
measuring the cool, neutral gas component
of galaxies, as well as a redshift indicator and 
kinematical probe.  The HI line can also be detected
in low mass systems with
little or no stellar emission. Blind HI
surveys thus yield samples which include a
population ``underappreciated'' in optical surveys.


A number of intergalactic HI clouds have been reported in the literature,
some in pairs and groups or in the vicinity of large disk galaxies 
(e.g. Schneider \etal 1983; Sancisi \etal 1987;  
Kilborn \etal 2000; Ryder \etal 2001),
others the result of disturbances
from interactions within a cluster
(Minchin \etal 2005, 2007; Oosterloo \& van Gorkom 2005; Haynes \etal 2007).
The serendipitous detection of HI1225+01 (Giovanelli \& Haynes 1989; 
Chengalur \etal 1995) revealed a binary structure, with the NE component
found to be associated with a dwarf galaxy, while the SW component
has failed to yield any 
optical light down to a surface magnitude 
limit of 27 mag arcsec$^{-2}$ (Salzer \etal 1991). 
A recently reported HI feature, VirgoHI21, 
was detected north of the spiral galaxy NGC 4254 (Minchin \etal 2005) and
interpreted as a fairly massive "dark galaxy". ALFALFA survey data, 
which provides full coverage of the region, indicate that VirgoHI21 is not 
an isolated dark galaxy, but rather a tidal tail of NGC 4254 resulting from
the high speed encounter of that galaxy with a now distant interloper 
(Haynes \etal 2007; Duc \& Bournaud 2008).

A galaxy traveling through a cluster can be affected by a 
variety of mechanisms, the most commonly invoked being ram pressure stripping. 
The study of HI deficiency in Virgo spirals was pioneered by Davies \& Lewis (1973),
with the first reliably measured deficiencies reported
by Chamaraux, Balkowski \& Gerard (1980). Giovanelli \& Haynes (1983) showed
that HI deficient galaxies in Virgo had truncated HI disks. These results
were extended to other clusters (Giovanelli \& Haynes 1985; Haynes \& Giovanelli
1986; Haynes, Giovanelli \& Chincarini 1984) and mapped with  higher resolution
(Cayatte \etal ~1990; Chung \etal ~2007).
Oosterloo \& van Gorkhom (2005) recently reported the discovery
of a HI cloud in the vicinity of the Virgo spiral 
NGC 4388 with a HI mass of $3.4~\times~10^8$\msun, resolved
into distinct clumps. They show a clear connection to NGC 4388.  
It has also been shown that galaxies  
can be ``harassed'' by the collective effect of the cluster potential and
multiple high speed encounters with other cluster members (e.g. Moore \etal 1996;
Mihos \etal 2005).
The hot intracluster gas is a hostile environment for the cold material
removed from cluster galaxies through these processes. The interaction
remnants represented by the latter are thus transient features. A full census
of their number and estimates of longevity can provide insights on the metal
enrichment of the intracluster gas.






Virgo is the nearest rich cluster of galaxies, thus providing the best
target for the detailed study of many environment driven effects on the 
evolution of galaxies. With over 1300 cataloged detections (Binggeli \etal 1985), 
the overdensity associated with the Virgo cluster extends over more than 300 
square degrees in the sky, distributed in
several subclumps.  The largest two clumps are centered around large elliptical galaxies, 
M87 for the main clump, and M49 for the southern subclump.  
The morphology--density relation is well established in the cluster, with late type 
objects thought to be still falling into the cluster, as it dynamically evolves.
The dynamically vulnerable, extended HI disks of late type galaxies can be easy
victims of the processes discussed above.
Optical data bases, such as the Digital Sky Survey (Lasker \etal 1990) and the Sloan Digital
Sky Survey (York \etal 2000), provide photometry and targeted spectroscopy of cluster 
objects, as does the compendium of Virgo galaxies data of GOLDMINE website 
(http://goldmine.mib.infn.it/ ; Gavazzi \etal 2003). Individual galaxy distances are growing
in number thanks to the ACS Virgo Cluster survey, which focuses on early-type galaxies 
via the surface-brightness fluctuation method (Mei \etal 2007).
Targeted HI surveys also have played an important role in the overall characterization
of the cluster (see, e.g. GOLDMINE for references).

The HI Parkes All-Sky Survey
(HIPASS; Barnes \etal 2001) has mapped 30000 deg$^{2}$, part of which covers
Virgo in the region 12$^{h}<$ R.A. $<13^{h}$ and +2$^{\circ} < $ Dec. $ < +20^{\circ}$.  HIPASS catalogs available
from NED list 125 galaxies in this region.
The Arecibo Legacy Fast ALFA Survey (ALFALFA; Giovanelli \etal 2005) is more  
sensitive and has significantly higher angular and spectral resolution than HIPASS.
It is thus providing a much more incisive look into the HI properties of the cluster.
While the HI content of many individual spiral galaxies has been mapped in detail 
with interferometry (Cayatte \etal 1990, 1994; Chung \etal 2007), no survey has 
covered the Virgo cluster region with the sensitivity and completeness of ALFALFA. 
It is a primary goal of the ALFALFA survey to study and characterize the local HI 
Universe, including the Virgo cluster.  
ALFALFA will provide a fair, well-sampled and homogeneous spectral dataset, covering 
7000 deg$^{2}$ of the
high galactic latitude sky out to a redshift of $cz_{\odot} \sim 18000 $~\kms.  
Approximately 250 deg$^{2}$
of the Virgo cluster in the range $11^h44^m <$ R.A.(J2000) $< 14^h00^m$ and 
$+8^\circ <$ Dec.(J2000) $<+16^\circ$ have been completely surveyed as of late 2007,  
with catalogs of extracted source parameters already released
for publication (Giovanelli \etal 2007; Kent \etal 2008).
A key feature of the survey is its sensitivity to low HI mass objects --
ALFALFA will provide a complete HI census down to $2 \times 10^7$\msun~ at the Virgo distance. 
Sampling the low-mass population will be key in determining the HI mass 
function in the cluster environment, as well as in the field.
The ALFALFA dataset in Virgo has already yielded many new interesting results 
(Kent \etal ~2007; Haynes \etal ~2007; Koopmann \etal ~2008). 

In this paper we give a detailed description
of a multi-component HI cloud complex in the Virgo
cluster region initially reported by Kent \etal (2007).  
In section 2 we describe the observations and data reduction method
within the context of ALFALFA and the cloud detections made with that dataset.  
In section 3 we detail
follow-up observations conducted with the Very Large Array.  In section 4
we discuss CO observations of the cloud.  Section 5 discusses the environment of the cloud complex.
Section 6 discusses possible formation mechanisms for the cloud complex.  Section 7
summarizes the results of the paper.
Throughout the paper we assume a distance to the Virgo Cluster of \dv=16.7 Mpc, and 
refer to the heliocentric reference frame for all velocities. 

\section{Discovery of a HI Cloud Complex}
\label{AO}

The ALFALFA observing strategy is described in detail by Giovanelli \etal (2005).  
The data are obtained
in a meridian transit drift mode with the Arecibo $L$-band Feed Array (ALFA).
Raw scans consist of 14 spectra (7 beams $\times$ 2 linear polarizations/beam), 
sampled at a rate of 1 Hz.
The spectra span a 100 MHz bandwidth with 4096 channels per polarization, centered at 1385 MHz.
The resulting spectral resolution is 24.4 kHz before smoothing, corresponding to $\delta V = 5.1$~\kms ~at the rest frequency of the 21cm HI line.
Scans are calibrated, baselined, and flagged for radio frequency interference, and then
combined into regularly sampled data cubes. 

Details of the ALFALFA data processing can be found in Kent (2008, PhD thesis).
Briefly, frequency channels in each grid have been baselined
with linear fits and then ``flatfielded'', where a median subtraction
has been performed in both the right ascension and declination directions.
A matched-filter algorithm was used for signal detection,
with manual follow-up (Saintonge 2007). The careful attention to signal 
extraction, analysis, and data quality has
proven to be invaluable in detecting faint objects 
and optimizing the output of the signal extraction process.

The cloud complex was discovered in the ALFALFA data obtained in the Spring 2005 campaign, 
which sampled the Virgo core region 
($12^{\rm{h}} < \alpha < 13^{\rm{h}}, 8^{\circ} < \delta < 16^{\circ}$). 
The data presented here are taken from a $2^\circ.4\times 2^\circ.4$ data cube centered 
at $\alpha= 12^{\rm{h}}20^{\rm{m}},\delta= 9^{\circ}00^{\prime}$ (J2000). The system 
temperatures of the ALFA receivers during the observations were in the range 
$26~\rm{K} < \rm{T}_{sys} < 30~\rm{K}$, yielding a root mean square (rms) noise of 
$\sigma_m=2.5\,$mJy/beam in channels with $\delta V=5.1$~\kms. The characteristics of the 
ALFALFA data cube from which the data are taken are summarized in Table~\ref{AOobs}. 


\subsection{ALFALFA Observations}
\label{AOdetect}

 The cloud complex is located near 
$\alpha\sim12^{\rm{h}}30^{\rm{m}}, \delta\sim9^{\circ}30^{\prime}$ -- this places it 
2.9$^{\circ}$~(845 kpc in projection) South of M87 and 1.5$^{\circ}$~(432 kpc in 
projection) north of M49. We discuss the optical environment of the cloud complex in 
Section 5.
 
The complex consists of five distinct emission features or ``clouds", which we denote C1--C5.
Together, they span approximately 35\arcmin~(170~kpc in projection at the cluster distance)
on the sky and 130 \kms~ in velocity. Channel maps of the ALFALFA dataset in the vicinity of
these detections are shown in Figure~\ref{AOcontours}, and total intensity (zeroth moment)
and intensity-weighted velocity maps of the region is in Figure~\ref{AOmainimage}. An
integrated spectral profile for each cloud is presented in Figure~\ref{AOspectra}. 

The individual properties of the clouds derived from the ALFALFA data are given in 
Table~\ref{AOparams}; all parameters are computed in the manner described by Giovanelli 
\etal (2007).  The spatial centroid of each cloud is in col.~(2). Its accuracy depends 
on the source strength, and varies from an average of $\sim15\arcsec$~for the brightest 
features to $\sim30\arcsec$~for the faintest ones. The heliocentric velocity $cz_\odot$, 
width at 50\% of the peak W50 and total flux $F_c$ of the integrated spectral profiles in 
Figure~\ref{AOspectra} are in cols.~(3)--(5). The signal-to-noise ratio $S/N$ of the 
detections is in col.~(6), and is given by
\be
        S/N=\Bigl({1000 \, F_c \over W_{50}}\Bigr){w^{1/2}_{smo}\over \sigma_{rms}} \;\;,
        \label{snr}
\ee
where $F_c$ is in Jy \kms, $W50$ is in \kms, $w_{smo}$ is a smoothing width equal to the number of $10\,$ \kms~bins bridging half the signal, and $\sigma_{rms}$ is the rms noise (in mJy) across the integrated spectrum at 10~\kms\ resolution. The HI mass \mhi~ for each cloud is in col.~(7), and is computed assuming that the clouds are optically thin and at the Virgo distance \dv=16.7 Mpc:
\be
M_{HI}/M_\odot = 2.356 \times 10^5 \, D_V^{2} \, F_c \;\;,
\label{HImass}
\ee        
where \dv\ is in Mpc and \fc\ is in Jy~\kms. The uncertainties on \mhi\ in Table~\ref{AOparams} and elsewhere do not include that in the distance adopted, which is poorly constrained due to the large peculiar velocities of objects near or within the cluster.



\subsection{Cloud Morphologies and Kinematics from ALFALFA data}
\label{AOclouds}

\textbf{Cloud C1:}  C1 is the main cloud in the complex and the highest $S/N$ detection 
in this region. It is marginally resolved by the ALFA beams (Figure~\ref{AOmainimage}). 
Its integrated profile is symmetric and narrow with a peak flux density of $38\,$mJy 
(Figure~\ref{AOspectra}). C1 has one of the largest HI masses in the complex at 
$M_{HI} = (1.62 \pm 0.04) \times 10^8$ \msun. There is a faint, uncataloged optical 
feature visible in SDSS images in the vicinity of C1; its relationship to the HI cloud 
is discussed in Section 5.

\textbf{Cloud C2:}  This isolated cloud is the faintest ALFALFA detection in the region 
at $\sn=6.5$, and it has the lowest \hi\ mass.  C2 is close to the ImV dwarf galaxy 
VCC~1357 (Binggeli \etal 1985), but the ALFALFA centroid is offset from the optical 
position of the latter by 2\arcmin\ to the West. No optical redshift is available for 
VCC~1357. An Arecibo single-beam observation centered on VCC~1357 with similar 
sensitivity to the ALFALFA data is presented by Hoffman \etal (1987). The properties 
of their detection are identical to those of C2 within the measurement uncertainties; 
these and the ALFALFA observations have likely uncovered the same object. However, 
the previous association of VCC~1357 with this \hi\ source is now in doubt by the 
evidence for an offset in position between the two, and by the detection of the other 
\hi\ complex clouds presented here. The positional offset between VCC~1357 and C2 is 
confirmed by the VLA data discussed in the next section. The possible relationship 
between C2 and VCC~1357 is discussed in Section 5.

\textbf{Cloud C3:}  This northernmost component of the complex is unresolved by the 
ALFA beam. Its integrated profile appears asymmetric with more emission on the 
high velocity side of the peak (Figure~\ref{AOspectra}), but this may be an artifact 
of the poor \sn. C3 is not connected to the main clouds C1 and C4 at the sensitivity
obtained. There is no discernible optical counterpart to C3 in the SDSS or DSS survey 
images of the region.

\begin{figure}
\begin{center}
\includegraphics[width=6.0in]{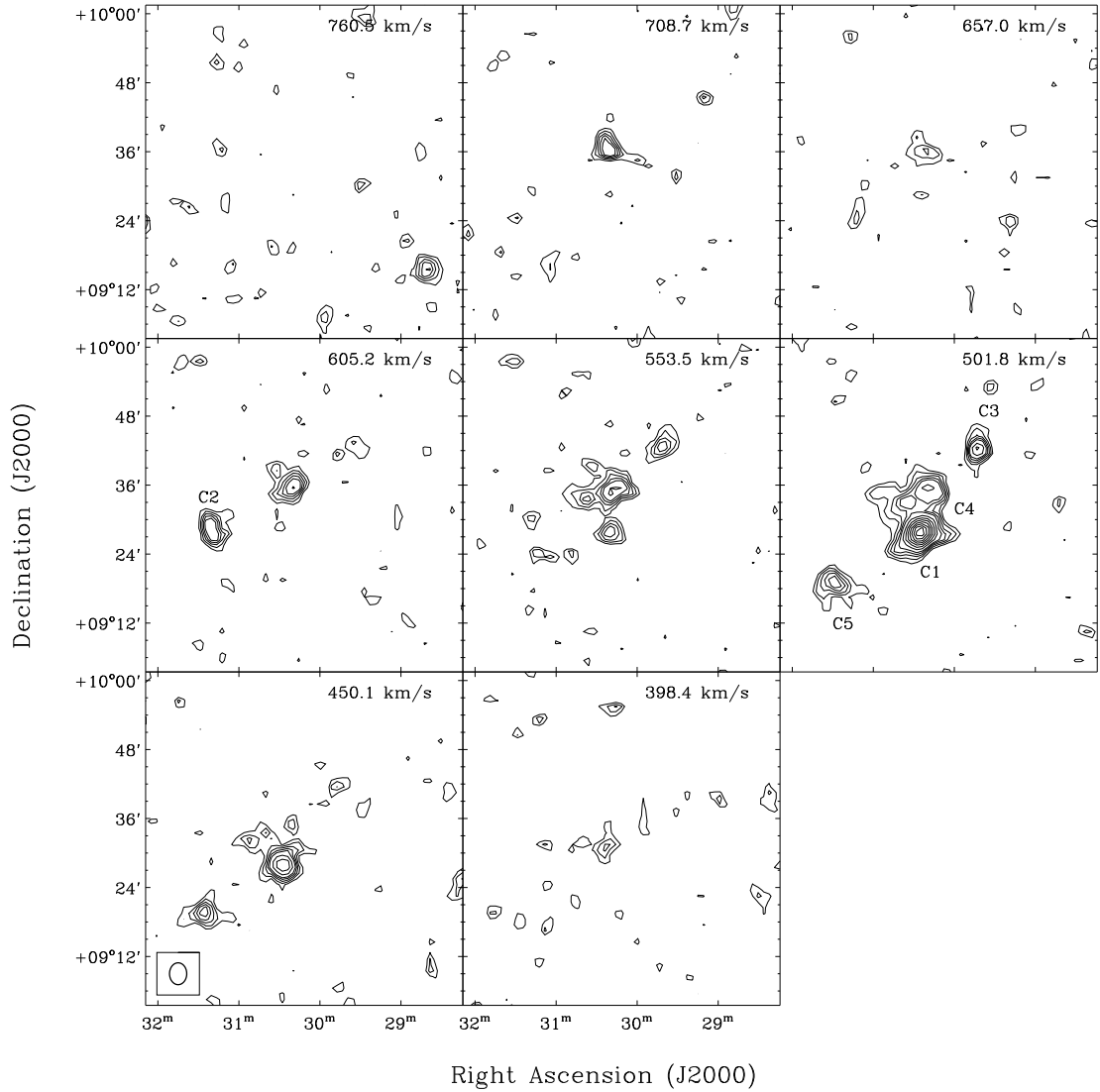}
\caption{Channel maps of the cloud complex from $760$~\kms~to~$400$~\kms in the ALFALFA survey data.  
          Contours are at (2, 3, 4, 5) and 6--26 mJy/beam in 2 mJy/beam intervals.  The velocities indicated 
         in the upper right corner
         represent the central channel velocity of the map.  The maps are boxcar smoothed at the indicated velocities
         +/- 5 channels (at $\sim5$~\kms~resolution).  The five HI cloud detections are indicated as C1 through C5.
          The ellipse in the lower left indicates the ALFALFA beam size ($3^{\prime}.3 \times 3^{\prime}.8$).
\label{AOcontours}}
\end{center} 
\end{figure}

\begin{figure} 
\begin{center} 
\includegraphics[width=7.0in]{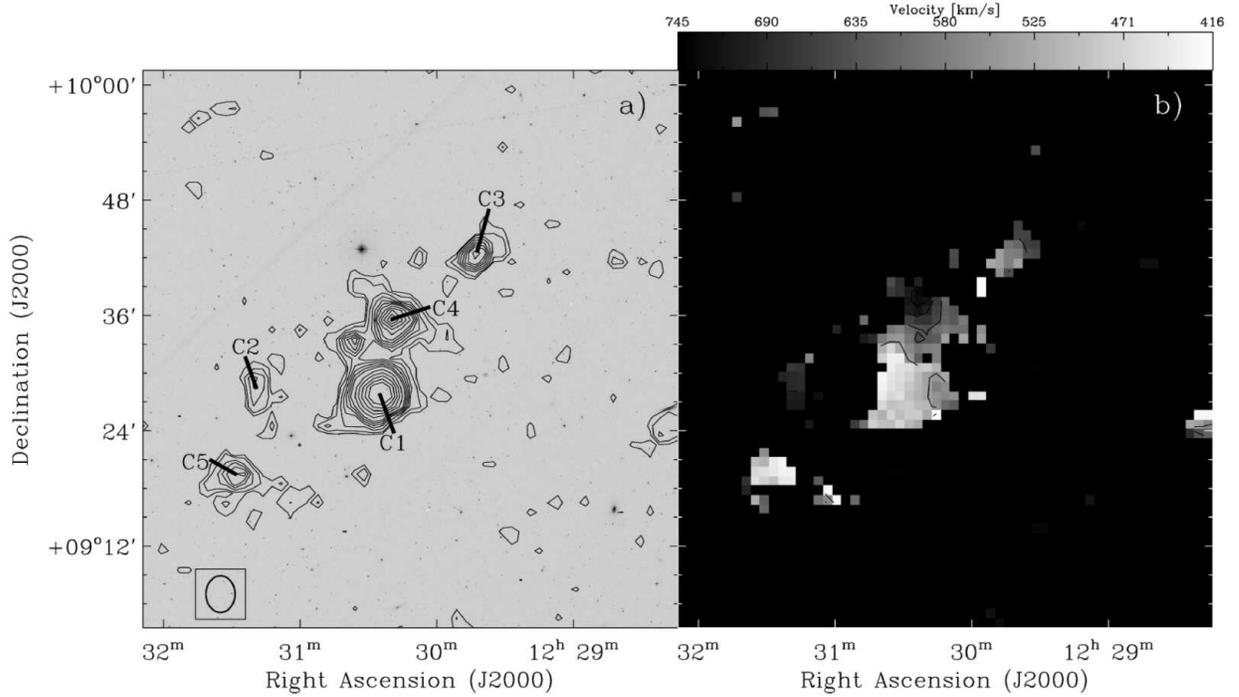} 
\caption{Global HI distribution and kinematics of the cloud complex.  $a)$
One square degree total intensity (zeroth moment) map of the cloud
complex field.  The cloud identifiers from Table~\ref{AOparams} are indicated.
The \hi~contours are 50, 100, 200, 300, 400, 500, 600, 800, 1000, 1200, 1400, 1600~mJy \kms~beam$^{-1}$.
The background image is taken from the 2nd generation Digital Sky Survey B-band plates (Lasker \etal 1990). 
The ellipse in the lower left indicates the ALFALFA beam size ($3^{\prime}.3 \times 3^{\prime}.8$).
$b)$ Intensity-weighted velocity (first moment) map.  The linear scale bar ranges from 416-745 \kms.
Contours indicate 450, 500, 550, 600, 650, 700 \kms.  A color version of this figure is available
in the electronic edition of the Journal.
\label{AOmainimage}} 
\end{center} 
\end{figure}

\begin{figure}
\begin{center}
\includegraphics[width=2.75in]{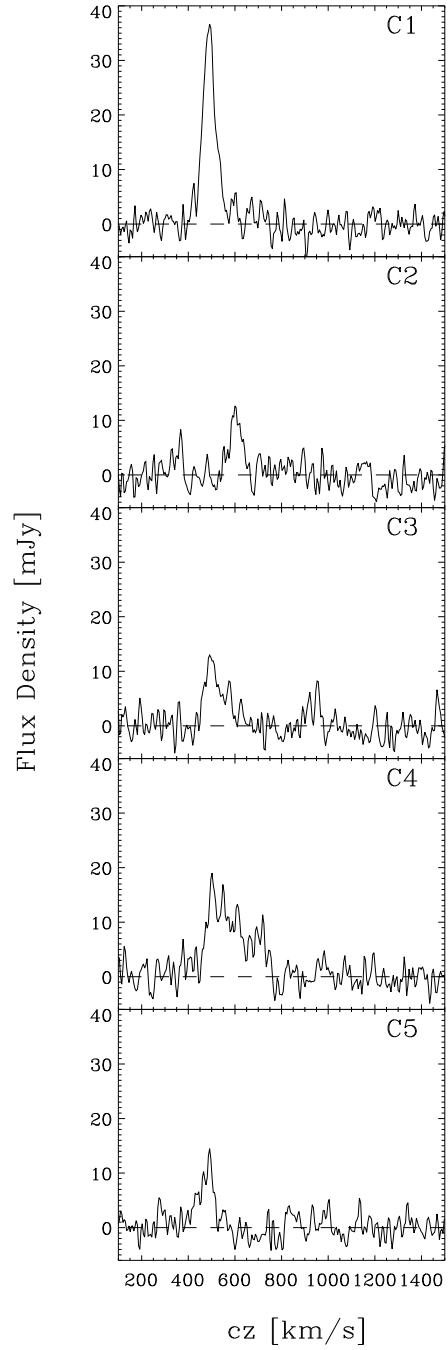}
\caption{Integrated spectral profiles for C1-C5 at a resolution of $\chan = 5.1~\kms$.  
The average RMS across the five spectra is 2.13 mJy/channel.
\label{AOspectra}}
\end{center}
\end{figure}

\textbf{Cloud C4:}  This cloud appears to be connected to the main cloud C1,
with its centroid located just 6$^{\prime}$.6 to the North of the latter (Figures~\ref{AOcontours}~and~\ref{AOmainimage}). Its \hi\ mass, $M_{HI}= (1.66 \pm 0.08) \times 10^8$  \msun, is comparable with that of C1 but its integrated profile is significantly 
broader.  The spectrum is asymmetric with more emission on the high-velocity side of 
the peak (Figure~\ref{AOspectra}), and the cloud is (poorly) resolved into a collection 
of smaller clumps by the ALFA beam (Figure~\ref{AOcontours}). Line broadening is thus
likely to arise from this superposition, rather than from coherent rotation. There is 
no discernible optical counterpart to C4 in the SDSS or DSS survey images of the region.

\textbf{Cloud C5:}  This cloud is located 17\arcmin.4\ Southeast of the main cloud C1. 
Its integrated profile appears to be symmetric, and implies an \hi\ mass that is a third 
that of C1 at $M_{HI} = (5.9 \pm 0.4) \times 10^7$ \msun. At the ALFALFA sensitivity, C5 
is not connected with any other complex clouds. There is no discernible optical 
counterpart to C5 in the SDSS or DSS survey images of the region.




\section{Aperture Synthesis Follow-up Observations}
\label{VLA}

Aperture synthesis observations of the central region of the cloud complex detected 
by ALFALFA were obtained with the Very Large Array\footnote{The VLA is a 
facility of National Radio Astronomy Observatory, which is operated by Associated Universities, Inc., under a cooperative agreement with the National Science Foundation.} (VLA) in two campaigns. On 2005 July 11, 5 hours on-source 
were obtained via rapid response observations 
in C configuration. On 2006 Jan 18 -- 23, dynamically scheduled observations during 
the D -- A configuration 
change yielded  $\sim1.5$ hours on-source 
with antennas that hadn't yet been moved. 
All observations had common pointing and spectral centers of
$\alpha=12^{\rm{h}}\,30^{\rm{m}}\,45^{\rm{s}},\, \delta= +9^\circ\,26'\, 0''$ (J2000) 
and $\nu=1417\,$MHz, 
and online Hanning smoothing was applied to yield 48.8 kHz channels over a total 
bandpass of 3.125 MHz. The 
observing setup was chosen to maximize the number of ALFALFA cloud detections 
accessible to the VLA in a single pointing: all of the clouds in Table~\ref{AOparams} 
except C3 fall within the resulting field-of-view and bandpass.

The data from the runs was reduced using the Astronomical Image Processing System 
({AIPS}; Greisen 2003). 
Standard flux, phase and 
bandpass calibration routines were applied. Continuum emission was removed from the 
data via linear fits to 
the average visibilities 
in line-free channels spanning 245 kHz at either end of the calibrated bandpass, 
yielding a net bandwidth of 
2.2 MHz sensitive to HI 
emission. 

 After calibration, the data from each run were combined into a single UV data cube and imaged using a variety of spatial 
and spectral weighting schemes. The dirty beam pattern was deconvolved from the data using 
the Multi-Scale Clean algorithm implemented in AIPS (Cornwell 2008; Greisen \etal, in prep.), in which components are extracted from a series of tapered images of the visibilities. We find that the resulting maps are not sensitive to the details of the deconvolution process, and therefore analyze the highest sensitivity, naturally-weighted cube with a synthesized beam width of 22\arcsec~(1.8 kpc at the Virgo distance).  
All maps and derived parameters are corrected for the attenuation of the 
primary beam, and averaged over 2 or 3 spectral channels to yield resolutions of $\delta V' =20.7$ \kms~
or $\delta V' = 31.0$~\kms, respectively. A summary of the aperture synthesis observing and map parameters is given in Table~\ref{VLAobs}. For clarity, all variables denoting parameters derived from the VLA observations are primed.

\subsection{\hi\ Aperture Synthesis Detections}
\label{VLAdetect}

 We make two detections in the VLA data, which correspond to the clouds C1 and C2 
identified in the ALFALFA survey data. Channel maps of these detections are shown 
in Figs.~\ref{C1chans} and~\ref{C2chans}. 
Contours are at multiples of the median rms map noise $\sigma_m'$ in the 
primary beam--corrected maps at that location, and negative contours are indicated by 
dashed lines. We find that the emission associated with C1 and C2 spans multiple 
synthesized beams over contiguous (but largely independent\footnote{The noise in 
each channel is weakly correlated by the continuum subtraction. We have verified 
that the details of this subtraction do not impact the detected cloud morphologies.})
 channels irrespective of the visibility weighting or image deconvolution scheme 
adopted: they are credible detections in the VLA data.

 Total intensity and intensity-weighted velocity maps for the clouds are shown in 
Figs.~\ref{C1moments} and \ref{C2moments}. The data cubes are blanked before these 
moments are computed: a mask is generated for each frequency channel by smoothing 
the data to half the angular resolution in Table~\ref{VLAobs} and blanking regions 
with fluxes less than 2\smp\ of that image. In addition, the intensity-weighted 
velocity maps in Figs.~\ref{C1moments}b and \ref{C2moments}b are computed only at 
locations with column density 
\nhp$\geq 10^{20}$ cm$^{-2}$.  Integrated spectral profiles for C1 and C2 from the 
VLA observations are in Figure~\ref{VLAspectra}. The $1\sigma$ error bars on each 
point in Fig.~\ref{VLAspectra} reflect \smp\ integrated over the emission region in that channel and a $5\%$
 calibration uncertainty. 

 The properties of C1 and C2 derived from the VLA data are given in 
Table~\ref{VLAparams}. Unless otherwise indicated, the parameters are computed 
in the same manner as their ALFALFA counterparts (see Section~\ref{AOdetect}). The 
location of the peak \nhp\ in the total intensity maps of Figs.~\ref{C1moments}a 
and \ref{C2moments}a is given in col.~(2). The centroid $cz_\odot'$ of the 
integrated profiles of Figure~\ref{VLAspectra} is in col.~(3), and $W50'$ of 
the profiles is in col.~(4). The values of $W50'$ are corrected for instrumental 
effects by assuming that the unbroadened profile is gaussian. The integrated flux 
density $F_c'$ and \hi\ mass $M_{HI}'$ are in cols.~(5)~and~(8), respectively.  
The maximum angular extent $a_{HI}'$ of each cloud is in col.~(6). We adopt the 
outermost locations where \nhp $= 10^{20}$ 
cm$^{-2}$ in Figs.~\ref{C1moments}a~and~\ref{C2moments}a as the cloud edges, and 
correct the measured values for beam smearing. The position angle \pahip\ at which \ahip\ is measured is in col.~(7). An estimate of the dynamical mass 
$M_{dyn}'$ of each cloud is in col.~(9), and is computed via:
 \be
 M_{dyn}' = (3.39 \times 10^4)\,a_{HI}' D_V \left( \frac{W50'}{2} \right)^2 \;\;,
 \label{Dynmass}
 \ee 
 where $a_{HI}'$ is the object diameter in arcminutes, $W50'$ is in \kms\ and the 
Virgo distance \dv\ is in Mpc. We note that $M_{dyn}'$ has physical meaning only 
if the clouds are self-gravitating and in dynamical equilibrium; these two assumptions 
may not be valid for C1 and C2 (see Section~\ref{discussion}).
 
 Position-velocity slices through the C1 and C2 datacubes are shown in Figs.~\ref{C1slice} and \ref{C2slice}, respectively. For C1 in Fig.\ref{C1slice}, one 3\arcsec-wide slice is oriented along \pahip, and the other is perpendicular to this axis. We note that the emission is unresolved both spatially and spectrally along the axis perpendicular to \pahip\ for C2, and we therefore show only the slice oriented along \pahip\ in Fig.~\ref{C2slice}.

\subsection{HI Morphologies and Kinematics of C1 and C2}
\label{VLAclouds}

 The VLA follow-up data provide important insight into the \hi\ morphologies and kinematics of C1 and C2.
 
\begin{figure}
\epsscale{1.0}
\plotone{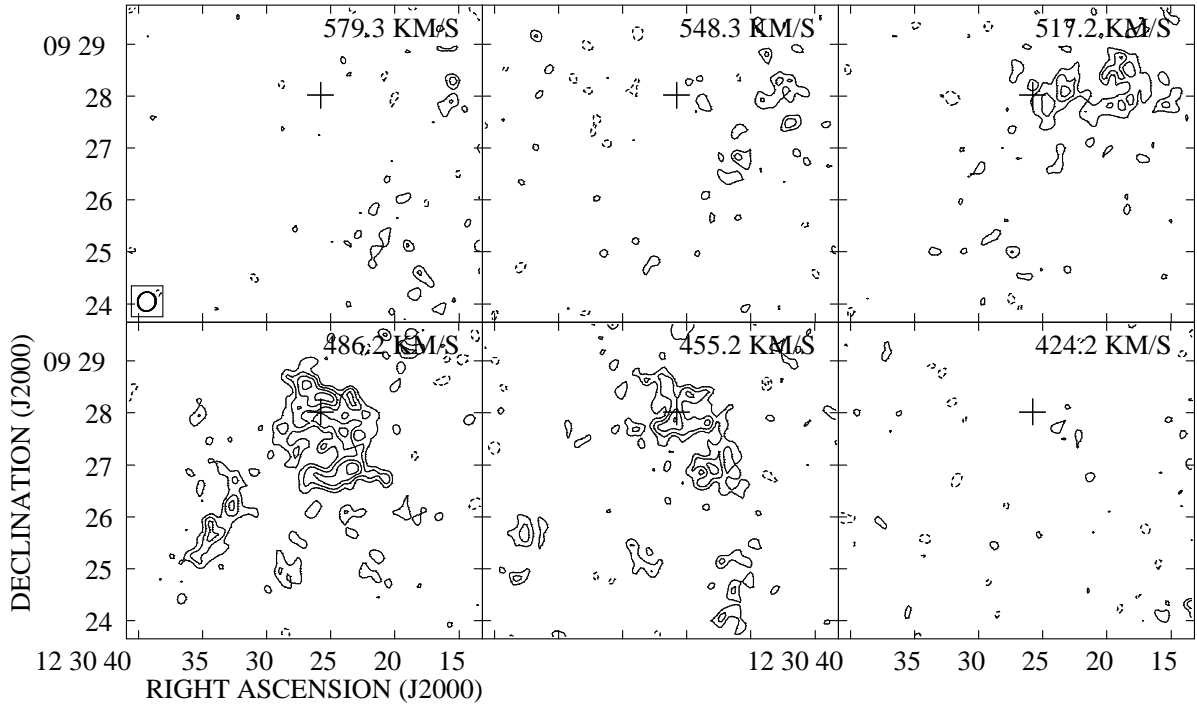}
\caption{Naturally-weighted channel maps for C1 from the VLA observations. The plotted channels are independent ($\chanp = 31~\kms$). Contours are at 0.35 $\times$ (-3, -2, 2 ($2\sigma_m^{\prime}$), 3, 4, 5, 6) mJy/beam; 
negative contours are represented with dashed lines. The cross denotes the C1 centroid in the ALFALFA data (Table~\ref{AOparams}). 
The heliocentric radial velocity is in the upper right corner of each panel, and the synthesized beam is in the lower left corner 
of the first panel.  A color version of this figure is available in the electronic edition of the Journal.
\label{C1chans}}
\end{figure}

\textbf{Cloud C1:}  Figs.~\ref{C1chans} and \ref{C1moments}a show that C1 has a 
disordered morphology at the resolution of the VLA observations, with the bulk 
of the emission stemming from an arc-like structure at $cz_{\odot}^{\prime}=486$~\kms. 
The apparent ``clumpiness" of the detected emission down to the synthesized beam 
width of $\sim22$\arcsec~(1.8~kpc) is not a deconvolution artifact, and suggests 
that the cloud exhibits structure on even smaller scales. The box in Fig.~\ref{C1moments}
a encloses a faint, uncatalogued optical feature that is coincident with a high $N_{HI}^{\prime}$ 
peak in C1; we discuss the implications of an association between these optical and HI 
sources in \S\ref{discussion}. Figs.~\ref{C1moments}b and \ref{C1slice} show that the cloud has no coherent
velocity structure. 

Figs~\ref{C1chans} and~\ref{C1moments} illustrate that there is very good agreement between the centroid 
of the C1 emission detected by ALFALFA and that detected by the VLA. The
global properties of C1 measured from the VLA data also correspond well with 
those obtained from the ALFALFA data, although less HI flux is detected in the
former ($F^{\prime}_{c}/F_{c} = 86 \pm$ 4\%, Tables 2 and 4). 
If this ``missing" flux is uniformly distributed over a circular region at least 1.5\arcmin~across, it would escape
detection at the 3$\sigma^{\prime}_m$ level in an optimally smoothed frequency channel of the VLA data. 
If it is contained in a coherent structure, we expect it to be
kinematically coincident with the arc in Fig. 6 since $cz_{\odot}-cz_{\odot}^{\prime}\sim$ 0.

\begin{figure}
\epsscale{1.0}
\plotone{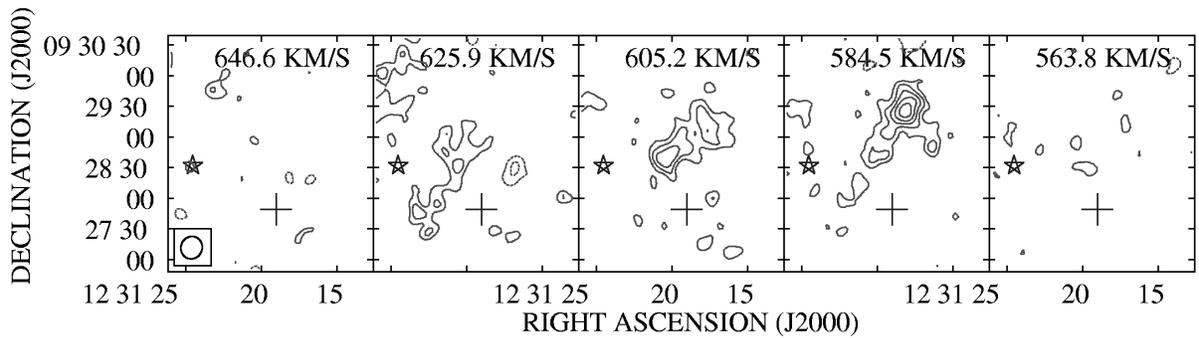}
\caption{Naturally-weighted channel maps for C2 from the VLA observations.  The plotted channels are independent ($\chanp = 20.7~\kms$). Contours are at 0.46 $\times$ (-3, -2, 2 ($2\sigma_m^{\prime}$), 3, 4, 5, 6) mJy/beam; 
negative contours are represented with dashed lines.  The cross denotes the C2 centroid in the ALFALFA data (Table~\ref{AOparams}), 
and the star denotes the optical position of VCC~1357 (Binggeli \etal 1985). The heliocentric radial velocity is in the upper right 
corner of each panel, and the  synthesized beam is in the lower left corner of the first panel. A color version of this figure is 
available in the electronic edition of the Journal.
\label{C2chans}}
\end{figure}
\begin{figure}
\epsscale{1.0}
\plotone{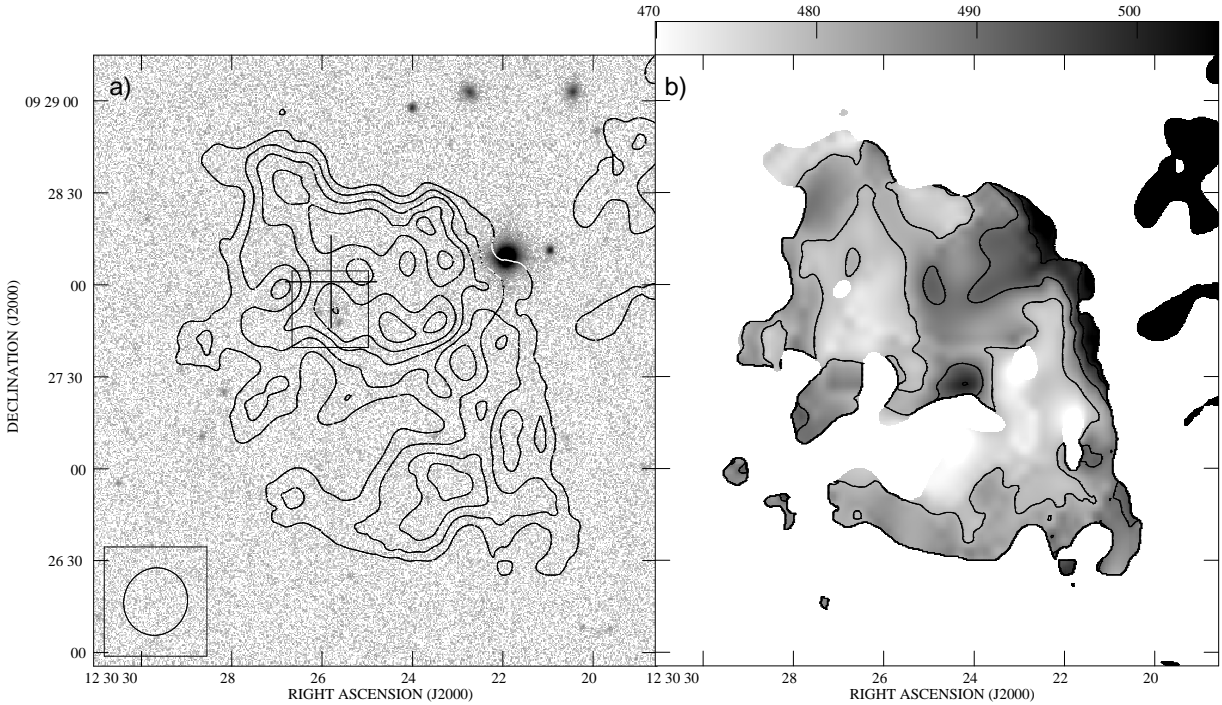}
\caption{\hi\ distribution and kinematics of C1 in the VLA data. $a)$ Total intensity map of C1 (contours) superimposed on an SDSS $g$ image (grayscale). Contours 
are at $\nhp=10^{20} \times\,$(1, 1.5, 2, 2.5, 3)~\cm, and the grayscale is plotted logarithmically. A very faint, uncatalogued source in the optical image is enclosed by the box. The cross indicates the centroid of the Arecibo detection for C1.  The synthesized beam is in the lower left corner of the panel. $b)$ Intensity-weighted velocity map of C1 in regions where $\nhp \geq 10^{20}~\cm$. The grayscale spans 470--505~\kms\ on a linear scale, as indicated by the wedge at the top of the plot. Contours are at (480, 490, 500, 510)~\kms.   A color version of this figure is available in the electronic edition of the Journal.
\label{C1moments}}
\end{figure}
\begin{figure}
\epsscale{1.0}
\plotone{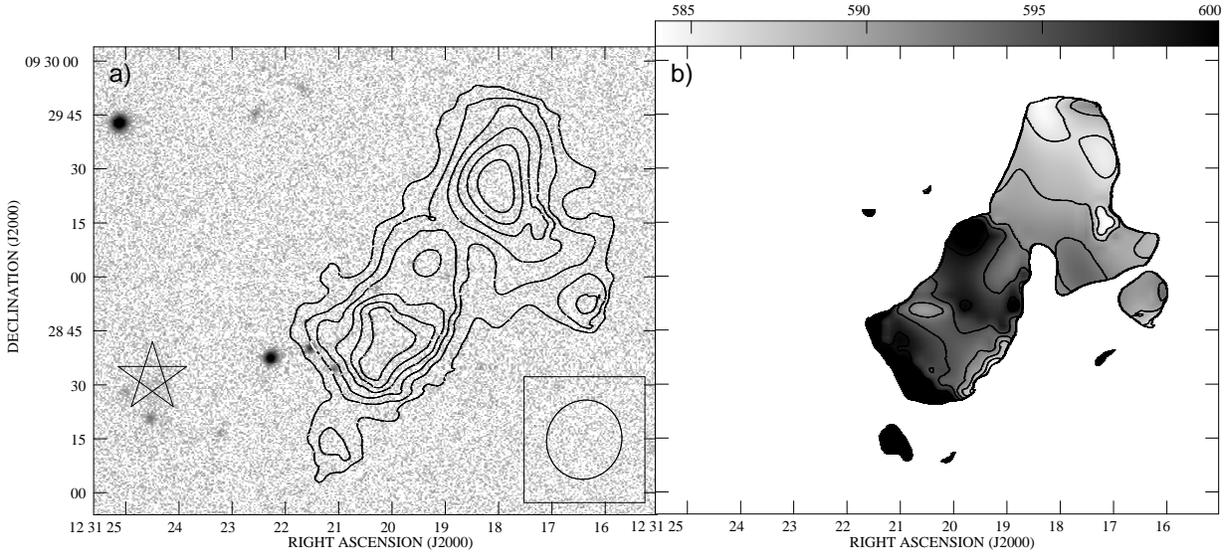}
\caption{\hi\ distribution and kinematics of C2 in the VLA data. $a)$ Total intensity map of C2 (contours) superimposed on an SDSS $g$ image (grayscale). Contours 
are at $\nhp = 10^{20} \times\,$(0.75, 1, 1.25, 1.5, 2, 2.25)~\cm, and the grayscale is plotted logarithmically. The star indicates the 
location of VCC~1357 (Binggeli \etal 1985); it is just visible in the optical image.  The synthesized beam is in the lower right corner of the panel. $b)$ Intensity-weighted velocity map of C2 in regions where $\nhp \geq 10^{20}~\cm$. The grayscale spans 585--600~\kms\ on a linear scale, as indicated by the wedge at the top of the plot. Contours are at (583, 586, 592, 595, 598)~\kms.   A color version of this figure is available in the electronic edition of the Journal.
\label{C2moments}}
\end{figure}
\begin{figure}
\epsscale{0.7}
\plotone{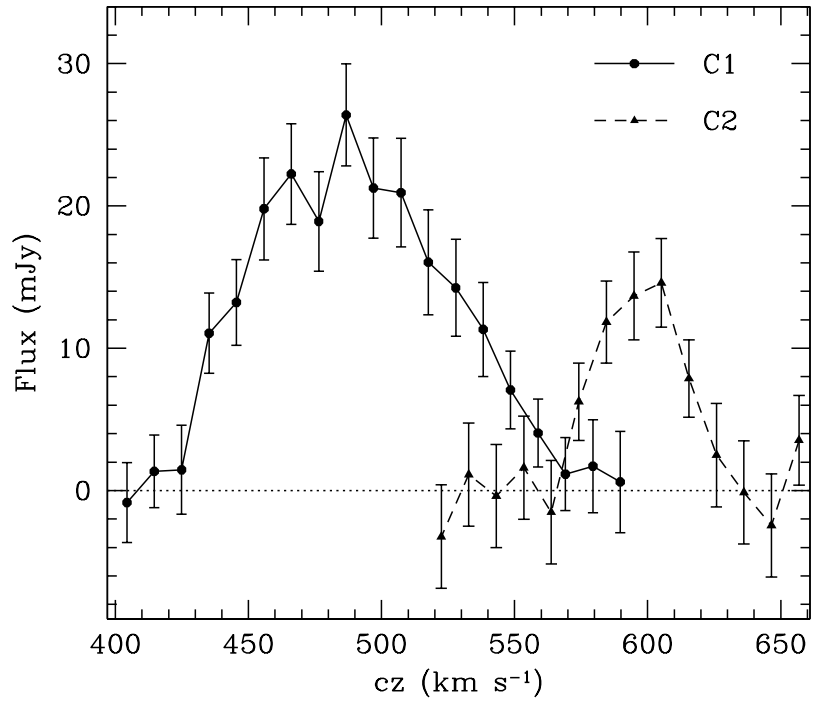}
\caption{Integrated spectral profiles of C1 ($\chanp = 31.0~\kms$; solid lines and circles) and C2 ($\chanp = 20.7~\kms$; dashed lines and triangles) from the VLA observations.  A color version of this figure is available in the electronic edition of the Journal. 
\label{VLAspectra}}
\end{figure}
\begin{figure}
\epsscale{0.8}
\plotone{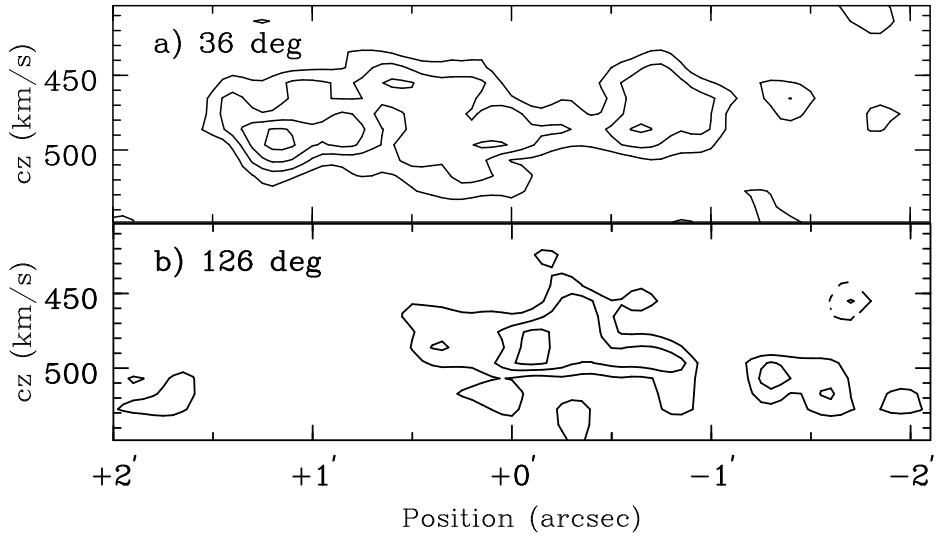}
\caption{Position-velocity plot of the \hi\ distribution in C1 along $a)$ $\pahip = 36$\arcdeg, the position angle at which the maximum linear extent \ahip\ was measured (see Table~\ref{VLAparams}), $b)$ the axis perpendicular to \pahip. In both panels, the position axis increases with increasing RA, and its origin corresponds to $ \alpha^{\prime} = 12^{\rm{h}}\,30^{\rm{m}}\,24.3^{\rm{s}},\, \delta^{\prime}= +9^{\circ}\,27'\, 41''$ (J2000). Contours are at 0.35 $\times$ (-3, -2, 2 ($2\sigma_m^{\prime}$), 3, 4, 5, 6) mJy/beam; negative contours are represented with dashed lines.
\label{C1slice}}
\end{figure}
\begin{figure}
\epsscale{0.8}
\plotone{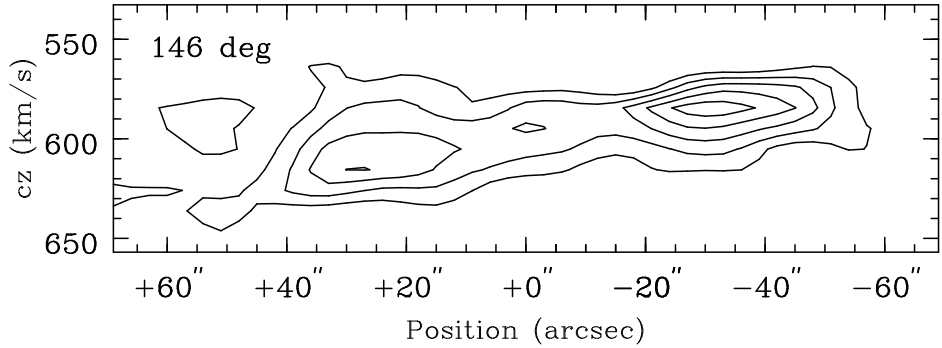}
\caption{Position-velocity plot of the \hi\ distribution in C2 along $\pahip = 146$\arcdeg, the position angle at which the maximum linear extent \ahip\ was measured (see Table~\ref{VLAparams}). The position axis increases with increasing RA, and its origin corresponds to $ \alpha^{\prime} = 12^{\rm{h}}\,31^{\rm{m}}\,19.2^{\rm{s}},\, \delta^{\prime}= +9^{\circ}\,29'\, 01''$ (J2000).  Contours are at 0.46 $\times$ (-3, -2, 2 ($2\sigma_m^{\prime}$), 3, 4, 5, 6) mJy/beam; negative contours are represented with dashed lines. 
\label{C2slice}}
\end{figure}

\textbf{Cloud C2:} Figs.~\ref{C2chans} and \ref{C2moments} illustrate that C2 
has a markedly different \hi\ morphology from C1. The emission 
in each channel of the VLA observations is barely resolved both spatially and 
spectrally, with the bulk of the emission in an unresolved feature  
at $ \alpha^{\prime} = 12^{\rm{h}}\,31^{\rm{m}}\,18.0^{\rm{s}},\, \delta^{\prime}= +9^{\circ}\,29'\, 30''$ (J2000) in the $cz^{\prime}_{\odot}=585~\kms$ channel. The total intensity 
map of C2 in Fig.~\ref{C2moments}a consists of two clumps: that to the Northwest 
results from the brightest \hi\ feature at $cz^{\prime}_{\odot}=585~\kms$, and 
that to the Southeast results from emission distributed over 
$585~$\kms $< cz_{\odot} <$ 625~\kms. Figs.~\ref{C2moments}b and \ref{C1slice} indicate that there 
is a $\sim15~\kms$ gradient in the velocity field  of C2. Given the poor 
sensitivity and resolution of the VLA data across the detection, it remains 
unclear whether or not this gradient stems from coherent internal motions.  

 A comparison between the integrated properties of C2 derived from the ALFALFA 
and VLA observations suggests that a small but statistically significant amount 
of \hi\ in this source has not been detected by the VLA 
($\mhi - \mhip = [1.0 \pm 0.5] \times 10^7 \msun$; Tables~\ref{AOparams}~and~
\ref{VLAparams}). There are few direct constraints on the morphology of the 
missing gas. However, given the gradient in the C2 velocity field (Fig.~
\ref{C2moments}b), the quantities $\czsun - \czsunp = (10 \pm 5)~\kms$ and 
$\wf - \wfp = (18 \pm 9)~\kms$ are consistent with a kinematically coherent 
extension to the Southeast beyond that detected at $\czsunp = 626~\kms$ in 
Fig~\ref{C2chans}. 
 
 The star in Figs.~\ref{C2chans}~and~\ref{C2moments}a denotes the optical 
location of VCC~1357 (Binggeli \etal 1985). The VLA observations therefore 
confirm the offset between this source and C2 suggested by the ALFALFA data 
(\S\ref{AOclouds}). It is thus clear that if C2 is associated with VCC~1357, 
then the gas in this system is displaced by $\sim2'$ (10~kpc in projection, 
or $\sim 10$ optical diameters; Binggeli \etal 1993) from the stars. We discuss 
the implications of an association between these \hi\ and optical features in 
Section 5.

\subsection{Non-Detections in the VLA Follow-up Data}
\label{VLAnondetect}

 While the complex clouds C4 and C5 fall within the field-of-view and bandpass 
of our VLA observations, they are not detected in the resulting dataset. This 
is not surprising given their projected distances of 11.2\arcmin\ and 12.5\arcmin\ 
from the pointing center in Table~\ref{VLAparams}. Given the integrated spectral  
profile shapes and centroids of these clouds in the ALFALFA survey data (Table~
\ref{AOparams}), we expect their emission to fall below the $3\smp$ detection 
limit in the VLA data cube if they are smoothly distributed over a circular 
region with a diameter $\gtrsim 1\arcmin$. The VLA non-detections therefore 
provide little insight into the morphologies and kinematics of C4 and C5.

\section {CO(3--2) Observations}

An exploratory observation of the CO(3--2) emission line consisting of a single pointing (R.A.(J2000) = 12$^{h}$30$^{m}$
27$^{s}$, Dec.(J2000) = +9$^{\circ}$28$^{\prime}$34.5$^{\prime \prime}$)
was carried out with
the APEX Telescope (G{\"u}sten \etal~2006) located in the Chajnantor Plateau
in northern Chile. This pointing was chosen because corresponds to the
peak of the HI emission in the VLA map. The observations took place during the night of January 1st, 2008, using the
the 345 GHz DSB heterodyne receiver APEX-2A. The beamsize (FWHM)
of the APEX 12-m telescope is 18$^{\prime \prime}$ at 345 GHz. The receiver
was tuned to CO(3--2) at $V_{\rm helio} =$ 490 km s$^{-1}$, with a total
bandwidth of 1550 km s$^{-1}$ and spectral resolution of 0.8 km s$^{-1}$.
Observations were performed with an atmospheric opacity at 220 GHz of 0.34,
with a mean PWV $\sim$ 2.1 mm yielding to a T$_{\rm sys}$(DSB) = 370 K.
Pointing using Saturn as reference was regularly performed and was better than 2$^{\prime \prime}.$ 
The total on-source integration time was 36 min, reaching a rms of
48 mK with a spectral resolution of 0.9  km s$^{-1}$. The antenna efficiency 
was estimated at 70\%.

Data reduction and analysis were performed using the CLASS package from
the GILDAS software (Pety 2005). Reduction consisted of baseline subtraction
and the integration of the individual scans. The adjacent channels were re-binned
to a velocity resolution of 22.9 km s$^{-1}$, leading to a rms of 12.9 mK in
$T_{mb}$. No clear detection was found.

These exploratory CO(3--2) observations allow us to place an upper limit
to the mass in molecular form in C1. In order 
to estimate a 3-$\sigma$ upper limit, we assume an expected line width
of 50 km s$^{-1}$, yielding to a $I_{\rm CO(3-2)} =$ 1.93 K km s$^{-1}$.
 We assume a conversion ``X'' factor between the CO(1--0) line and
the molecular hydrogen column density $N_{\rm H_{2}}$
of 2.8$\times$10$^{20}$ cm$^{-2}$ (km s$^{-1}$)$^{-1}$ (Bloemen \etal~1986).
We also
adopted a CO(3--2)/CO(1--0) integrated line intensity ratio of 0.4
which is the typical value of the Galactic disk (Sanders \etal~1993). These
values yield a H$_{2}$ column density limit for the observation of
$N_{\rm H_{2}} <$ 1.3$\times$10$^{21}$ cm$^{-2}$, averaged over a beam of 18$^{\prime \prime}$.
At a distance of 16.7 Mpc, this column density over a 1\arcmin~diameter is equivalent
to a molecular hydrogen upper mass limit of $M_{\rm H_{2}} <$ 4.0 $\times$10$^{8}$ M$_{\odot}$,
which is comparable to the HI mass (see Table 2.). Allowing for uncertainties
in the value of the X factor, the metallicity of the cloud gas, and the
CO(3--2)/CO(1--0) integrated line ratio, we can conclude that the gas mass
in the cloud C1 is unlikely to be dominated by the molecular component.

%
%

\section {The Environment of the Cloud Complex}\label{optenv}

The optical feature of unknown redshift appearing in the field of C1 and
discussed in Section 3.2 appears to be just above the SDSS $g$--band detection limit.
If the feature is approximately $\sim 10^{\prime\prime}$ across, then its $g$--band 
luminosity at the Virgo distance would be $L_{g}\sim 10^{6}~L_{\odot}$.
The stellar M/L ratios models of Bell \etal (2003) give $M^*/L^*\sim$1.6. 
If this feature is the optical counterpart of C1, then the
stellar to HI mass ratio is less than 0.01 and $M_{HI}/L_g> 170$.
An analogous exercise for C2 and the putative counterpart VCC 1357 yields 
a stellar to HI mass ratio of less than 0.05.

As for the possible optical counterparts of clouds C1 and C2, attempts to 
obtain optical redshifts were made in two occasions with the Palomar 5m 
Hale telescope. The first time was unsuccessful due to limited sky transparency,
the second due to lack of any discernible H$_\alpha$ emission or of any other 
measurable spectral signature. The possibility of an association between HI and 
optical features remains open.

\begin{figure}
\begin{center}
\includegraphics[width=2.0in]{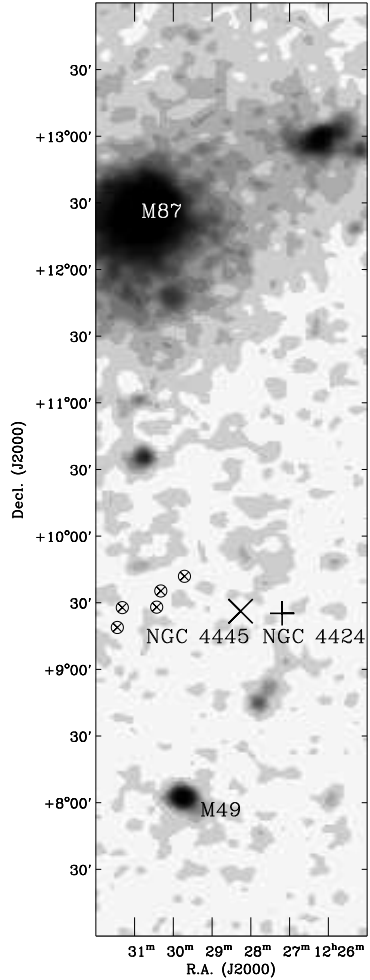}
\caption{The environment of the \hi\ cloud complex within the greater cluster area. The crossed circles indicate the five components
          of the complex discussed in this paper.  The large plus(+) indicates the 
          position of SBa galaxy NGC 4424 ($cz_{\odot}$ = 476~\kms).  The large X indicates the position 
          of Sab galaxy NGC 4445 ($cz_{\odot}$ = 354~\kms.  The X-ray peaks are labeled
          indicating Virgo cluster galaxies M49 and M87 (Snowden \etal~1995).  The symbols are not indicative of source sizes and 
          are shown only for positional indication.
\label{virgoenviron}}
\end{center}
\end{figure}

Figure \ref{virgoenviron} shows the Virgo Cluster X-ray emission in the vicinity 
of the HI cloud complex (Snowden \etal 1995). The large X--ray emission regions are 
centered on M49 and M87, with the HI cloud complex indicated with the crossed circles.
Near the projected location of the cloud complex, Vollmer \etal (2001) estimate that the
hot intracluster gas (ICM) density is $n_{icm} \sim 10^{-4}$ cm$^{-3}$, while 
Shibata \etal~(2001) have measured an ICM temperature
of $T_8=0.235$ in units of $10^8$ K.

\begin{figure}
\begin{center}
\includegraphics[width=6.0in]{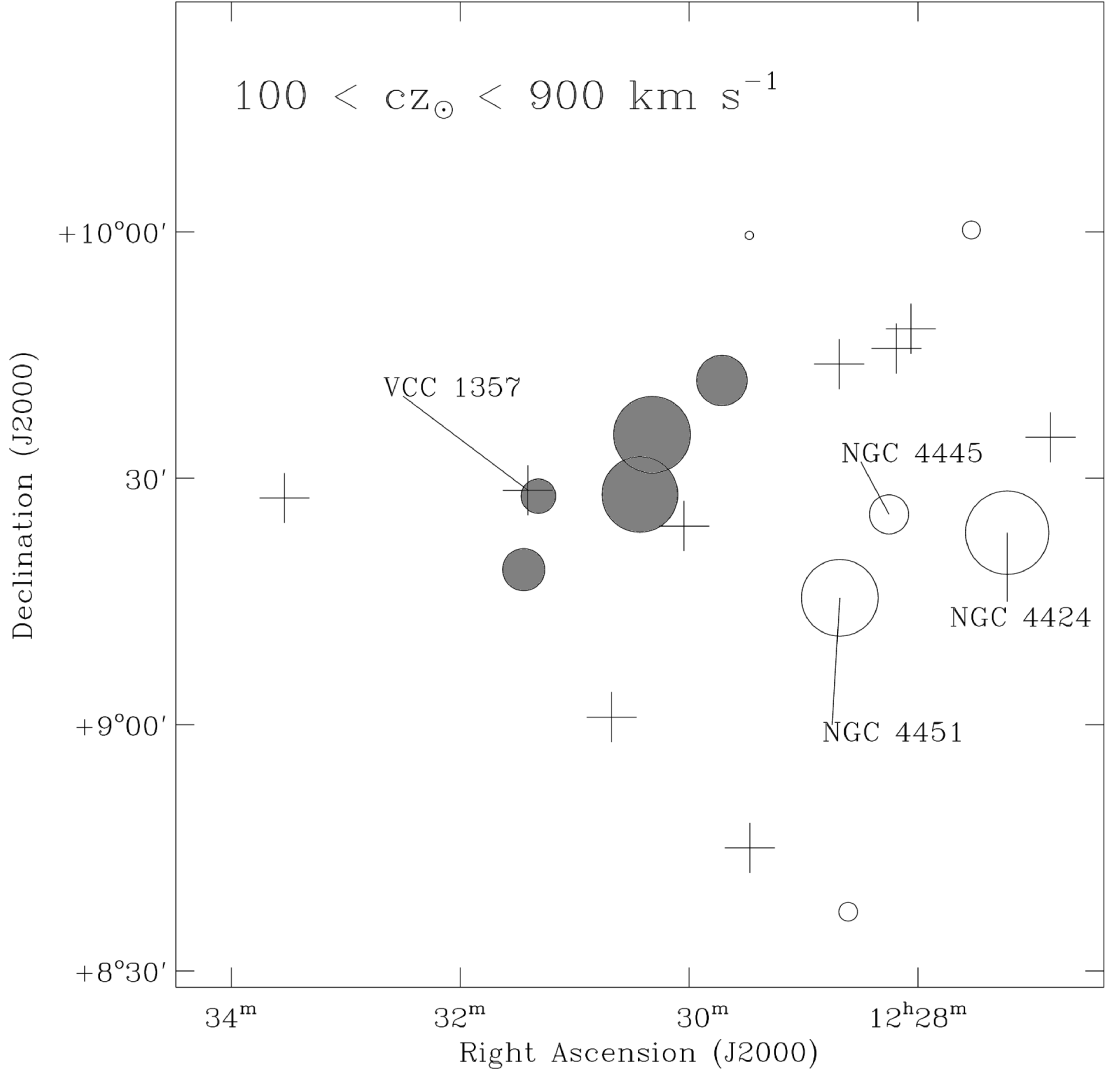}
\caption{Galaxies in the vicinity of the cloud complex.
         The plot shows all objects with 100 $<~cz_{\odot}~<$ 900~\kms~
         and within 1$^{\circ}$ of C1
         listed in the Arecibo General Catalog.  The cloud identifiers
         from Table~\ref{AOparams} are indicated by the shaded circles.
         The crosses indicate objects without HI measurements 
         from the ALFALFA survey.  Objects with HI masses as measured from ALFALFA
         are shown as circles, with the circle radius indicative of the HI mass. The smallest circle indicates
         log$_{10}$(M$_{HI}/$\msun)=7.30 and the largest circle indicates log$_{10}$(M$_{HI}/$\msun)=8.28.
         Galaxies pertaining to the possible origin of the HI cloud complex are labeled for reference.\label{skyarea}}
\end{center}
\end{figure}

Several galaxies are projected in the vicinity of the cloud complex. Figure~\ref{skyarea}
shows all cataloged objects within $1^{\circ}$ of C1 and with heliocentric velocities
between 100 and 900 \kms: 
galaxies detected in the 21cm line are plotted with a circle of area proportional to the
HI mass;  galaxies not detected in ALFALFA with redshift information from
other sources are plotted with crosses, most of which are early type systems.  
It is assumed that all galaxies are located at the distance of the Virgo cluster. 
Note that the velocities of the HI clouds range between 480 and 607 \kms, i.e. if they
are part of the Virgo cluster their line of sight velocity with respect to the cluster
reference frame is directed toward us, as the heliocentric velocity of the cluster is 
1150 \kms ~(Huchra 1988). Under the assumption that the clouds originated in the disk 
of a galaxy moving at high speed through the cluster, the velocity of the parent galaxy 
should also be incoming in the cluster reference frame, and most likely doing so at a 
larger velocity than the clouds themselves, as the latter would be decelerated by ram 
pressure after stripping. In the heliocentric reference frame, the parent galaxy should 
then have a {\it lower} velocity than those in the cloud complex. Two objects satisfy 
that condition: NGC 4424 and NGC 4445. Their locations are indicated by a plus sign 
(NGC 4424) and X symbol (NGC 4445) in Figure~\ref{virgoenviron}. In addition, as  we
discuss below, they are also extremely gas deficient for their size, which is evidence 
that they have been stripped within the last passage through the cluster. 

The SBa galaxy NGC 4424 (also known as UGC 7561 and VCC 9079) is located at 
$\alpha=12^{\rm{h}}27^{\rm{m}}11.6^{\rm{s}}, \delta=+09^{\circ}25^{\prime}14^{\prime\prime}$~(J2000).  
This places NGC 4424 at a projected distance of 236 kpc from the cloud center, 
and 900 kpc from M87. This peculiar galaxy is at a velocity ($cz_{\odot}$=437 \kms), 
near that of the main cloud. The angular size of the stellar component, 
$3^\prime.6 \times 1^\prime.8$, translates to a linear size of $17\times 9$ kpc 
at the Virgo distance. A detailed study by Cortes \etal (2006) characterizes it 
as having disturbed morphology; they also suggest that CO observations are 
indicative of non--circular gas motions and discuss convincingly the vulnerability 
of the galaxy to ram pressure stripping (see also  Kenney \etal~1996). NGC 4424 has 
a HI mass of 2.5$\times 10^8$ \msun~and estimations of HI deficiency values have 
been reported between of 0.75 (Chung \etal ~2007) and 1.09 (Helou \etal~1984). 
Those figures indicate that NGC~4424 has lost between 80\% and 90\% of its HI gas.
Chung \etal (2007) noted that NGC 4424 exhibits a one--sided tail extending to 
the southeast, 
which is also detected by ALFALFA, and that the HI disk is truncated at a radius 
smaller than that of the stellar component of the galaxy.  The direction
of the tail shows the southeast to northwest direction of motion
for NGC 4424; this motion is not compatible with the position of the cloud complex.  

NGC 4445 (also known as UGC 7587 and VCC 1086) is an edge on spiral, possibly of type
Sab, located at $\alpha=12^{\rm{h}}28^{\rm{m}}15.9^{\rm{s}}, \delta=+09^{\circ}26^{\prime}11^{\prime\prime}$~(J2000).
Its heliocentric radial velocity is $cz_{\odot}$=354 \kms~and the HI mass 
$5.5\times 10^7$ \msun ~(Kent \etal ~2008). Solanes \etal ~(2002) estimate its
HI deficiency between 0.98 and 1.11, i.e. NGC 4445 has lost about between 90\% 
and 93\% of its HI gas. With a major axis angular diameter of $2^\prime .6$, its
linear size is about 13 kpc. We are not aware of HI synthesis data of this
object; none are found in the VLA data archive.

\section{Discussion}
\label{discussion}


One possibility that deems consideration is that the clouds are gravitationally
bound structures, embedded in their own dark matter halos. 
If an HI cloud is gravitationally bound, its total dynamical mass within the
HI radius can be computed from Equation 3, assuming spherical symmetry.  
Such mass is 2.3$\times 10^9$ \msun~for C1 and 0.3$\times 10^9$ \msun~for C2 as 
calculated from the aperture synthesis measurements.  
The density of such objects would be $1.4\times 10^{-25}~\rm{g~cm^{-3}}$~for C1 
and $1.0\times 10^{-25}~\rm{g~cm^{-3}}$~for C2. These densities can be 
compared to the critical density, $\rho_{crit}=\left(1.9 \times 10^{-29} ~\rm{g~cm^{-3}}\right)h^2$,
where $h=0.7$. Were dark matter to provide the gravitational binding and to extend
well beyond the visible, baryonic component, the total halo mass of, e.g., C1 would exceed
$10^{10}$ \msun. In that case, it would be surprising that the VLA data
do not exhibit any trace of ordered motions in the gas, nor does the visible
matter show any degree of central concentration, thus violating the mass--concentration
relation for halos. A similar issue can be raised for the whole cloud complex.
If bound by self--gravity, its total dynamical mass would be of order of $10^{11}$~ 
\msun~and its mean density $2.1\times 10^{-27}~\rm{g~cm^{-3}}$. Again, the cloud complex 
shows no ordered motions. The possibility that the clouds in the complex are a group of 
optically dark galaxies embedded in their own dark matter halo appears relatively
implausible.

For a purely gravitational event, we may expect stream-like topology in the removed gas, 
which is not observed. 
High speed galaxy-galaxy encounters tend to produce relatively small damage in terms of gas mass loss 
by the victim, albeit spectacular in spatial extent as in the case of NGC 4254/VirgoHI21. For 
NGC~4424 and NGC~4445, the damage would have been quite substantial (more HI is seen in the
cloud complex than in either of the galaxies themselves) and the topology of the remnant 
is not particularly stream-like. These are not however very strong arguments and the
possible origin through a high speed close encounter cannot be excluded.

Ram pressure stripping is very likely if the galaxy's orbit takes it deep in the 
cluster potential well, and can result in severe gas depletion
from its outer disk (Cortes \etal 2006). 
With a radial velocity differing from the systemic velocity 
of the cluster by 700 \kms~and a projected location half way between M87 and M49, 
the likelihood that ram pressure is currently very effective is marginal; it could
however have been far more effective a few 10$^8$ years ago, if the orbit dipped closer
to the cluster center, independently on whether or not an encounter or merger took place.
As we will see, that likelihood is even higher for NGC 4445. 

\subsection{Orbital Motion Through the ICM: a Simple Simulation}

\begin{figure}
\begin{center}
\includegraphics[width=3.1in]{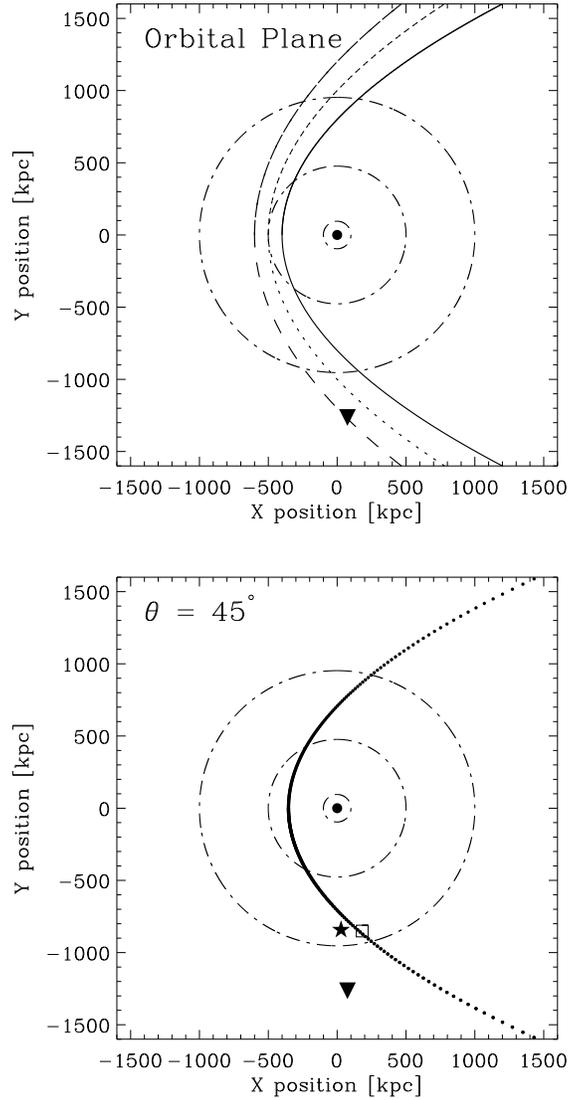}
\caption{
Simple schematic showing parabolic trajectories around the cluster.  The axes
are centered on the position of M87.  The top plot shows parabolic trajectories in the orbital plane
with $r_{p}$=400 (solid), 500 (dotted), and 600 (dashed) kpc.  The position
of M49 is shown as an inverted triangle.  The dot-dashed concentric circles
show radial distances of 100, 500 and 1000 kpc.  The bottom plot
shows the parabolic trajectory for $r_{p}$=500 kpc, corrected for an inclination 
of $\theta$ = 45$^{\circ}$ to the line of sight.  The positions of NGC 4445 (square) and HI cloud C1 (star) are also shown.
\label{orbit}}
\end{center}
\end{figure}

We consider the motion
of a galaxy as it dips into the ICM on a parabolic orbit of pericluster distance
$r_p$. Parametrized by the azimuthal angle $\phi$, as illustrated in Figure \ref{orbit},
the distance of the galaxy from the focal point (M87) is 
\be
r=\frac{2r_p}{1+\cos{\phi}}
\ee
while the orbital escape velocity is
\be
v_g=\sqrt{\frac{2GM_{clu}}{r}}
\ee
where $M_{clu}$ is the cluster mass. For simplicity, we assume the cluster mass is
concentrated in M87 and equal to $10^{14}$ \msun, rather than integrating the
motion for a changing cluster mass within the variable $r$, which is sufficient 
for our purposes. For the ICM density $n_{icm}$ we assume the
$\beta$--model (Schindler, Binggeli \& Boehringer 1999; Vollmer \etal
~2001), given by
\be
n_{icm}=n_{0} (1+ r^2/r^2_{clu})^{-3\beta/2} \label{beta}
\ee
with a central density of $n_{0}=0.04$ cm$^{-3}$ and a core radius of $r_{clu}=21$ kpc,
appropriate for the SW sector of the cluster and $r_{clu}=13.4$ kpc for the cluster as a
whole.

\begin{figure}
\begin{center}
\includegraphics[width=2.5in]{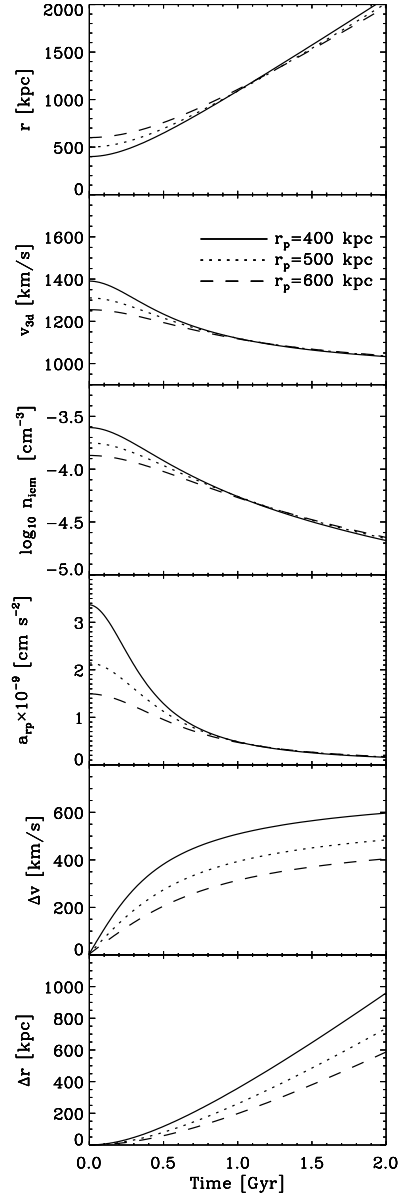}
\caption{\footnotesize{Results of a simple simulation of a galaxy moving in a
parabolic orbit through the ICM.  A cloud is detached
from the galaxy at the pericluster distance, and experiences
the dynamic effects due to ram pressure.  The plotted parameters
are not corrected for the inclination of the orbit to the line of sight.  Three
cases are shown for a pericluster distance
of $r_{p}$=400 (solid), 500 (dotted), and 600 (dashed) kpc.  Each plot
is depicted as a function of time in gigayears since pericluster passage.  The plots show (from
top to bottom), $r$, the distance from the center
of the cluster, $v_{3d}$, the velocity of the galaxy,
log$_{10}~n_{icm}$, the base 10 logarithm of the ICM density as the galaxy
 moves through the cluster, $a_{rp}$, the acceleration
on the detached HI cloud due to ram pressure, $\Delta v$, the
velocity accumulated by the cloud due to acceleration by ram pressure, 
and $\Delta r$, the separation between the galaxy
and cloud as they move in the parabolic orbit.}\label{kinematics}
}
\end{center}
\end{figure}

As the galaxy travels through the ICM, we assume the galactic
gas is stripped at pericluster; from there on, we estimate the ram acceleration
and integrate the growing separation  $\Delta r$ and $\Delta v$ between the galaxy
and clouds. Figure~\ref{kinematics} shows the orbital parameters for pericluster distances
of $r_{p}$ = 400, 500, and 600 kpc.  The best orbit in this simple exercise is
fit with $r_{p} \sim $ 500 kpc.  In that scenario, a galaxy reaches
a distance $r$ = 1100 kpc, 1.0 Gyr after pericluster passage.  The velocity
of the galaxy at $r$ = 1100 kpc is 1100~\kms.  Velocity ($\Delta v$) and
spatial ($\Delta r$) separation values of 200 kpc and 300~\kms~respectively are obtained
from Figure~\ref{kinematics}.  Corrections for the inclination along
the line-of-sight for $\theta \sim $ 45$^{\circ}$ give values
of $\Delta r \sim$ 140 kpc and $\Delta v \sim$ 200~\kms.  These numbers are in general
agreement with the observed values.

\subsection{Possible Formation Mechanisms}

We can now consider two possible scenarios of cloud formation as the galaxy moves through
the cluster:
(a) removal by ram pressure and (b) a high speed tidal interaction, of the type
responsible for VirgoHI21 (Duc \& Bournaud 2008; Haynes \etal 2007). 

Ram pressure stripping will occur
when the ram pressure arising from the motion of the galaxy at velocity
$v_{gal,3d}$ with respect to the intracluster medium of number density $n_{icm}$,
$n_{icm} m_p v_{gal,3d}^{2}$ (where $m_p$ is the proton mass), exceeds the 
restoring gravitational force that binds the
gas to the galaxy: $2\pi G \Sigma_{ism} \Sigma_d$, where $\Sigma_{ism}$ is the
gas surface density of the galactic disk and $\Sigma_{d}$ is the total 
surface mass density of the disk. This condition can be rewritten as
\be
\Bigl({n_{icm}\over 10^{-3}\, \rm{cm}^{-3}}\Bigr) 
\Bigl({v_{gal,3d}\over 10^3\, \rm{km~ s}^{-1}}\Bigr)^{2} \geq
11 \Bigl({M_d\over 10^{11} M_\odot}\Bigr)^2
   \Bigl({R_{gal}\over 10\, \rm{kpc}}\Bigr)^{-4}
   \Bigr({M_{ism}\over 0.1M_d}\Bigr) \label{ramp}
\ee
where $M_d$ and $M_{ism}$ are now the total disk and gas masses within the
radius $R_{gal}$. 

At the location of the cloud complex, some 845 kpc from M87, the density of the
intracluster gas (ICM) is significantly lower than near the center of the cluster.
For a $\beta$--model (Eqn.~\ref{beta}) at the present projected distance of the clouds from M87, $n_{icm} \sim 10^{-4}$
cm$^{-3}$.
In the case of NGC 4424, $v_{gal,3d}\simeq 700/ \cos{\theta}$ 
\kms, where $\theta$ is the angle between the line of sight and the galaxy 
velocity vector. We estimate a disk mass enclosed within $R_{gal}\simeq 10$ kpc 
of about $10^{10}$ \msun (a generous estimate, since the total velocity width of 
the HI emission of the galaxy is 61 \kms ~(Kent \etal ~2008) and the inclination 
of the disk, albeit unknown, is unlikely to be smaller than 45$^\circ$) and 
assume $M_{ism}/0.1M_d\sim 1$.
With those numbers, the ram pressure term roughly matches the restoring 
force at the larger galactocentric radii. Chung \etal (2007) also compute ram pressure 
effects in the range between 125 $< n_{icm} v_{gal}^{2} <$ 175 cm$^{-3}$ (\kms)$^2$  
and a restoring force range between 40 $< \Sigma_{ism}V_{rot}^2/R_{gal} <$ 475 
cm$^{-3}$ (\kms)$^2$ for NGC 4424.

In the case of NGC 4445, $v_{gal,3d}\simeq 800/ \cos{\theta}$ \kms. 
The radial extent of the HI disk is currently unknown, but we can estimate
the ram pressure and the restoring force at $R_{gal}\simeq 6$ kpc, as
well as its disk mass from a rotational velocity of 107 \kms, from the measured velocity 
full width of the HI line of 213 \kms.  Again, the ram pressure term is comparable
with the restoring force at the galactocentric radii at which most of 
the HI would be expected to be found in an unperturbed disk.

It is interesting to point out that NGC 4522, a very clear case of ram pressure 
stripping carefully studied by Kenney \etal ~(2004),
is projected on the same region of the cluster (RA=12$^h$ 33$^m$ 39.7$^s$, 
Dec=09$^\circ$ 10\arcmin 30\arcsec) albeit in a different velocity regime ($cz=2307$ \kms).
At the present location, Kenney \etal ~argue that the ICM density as described
by the smooth $\beta$--model of eqn. \ref{beta} is likely to be inadequate to
explain the stripping. They invoke the possibility that the ICM may be locally
denser than implied by the model.

\subsection {Timing Considerations}

\begin{figure}
\begin{center}
\includegraphics[width=3.0in]{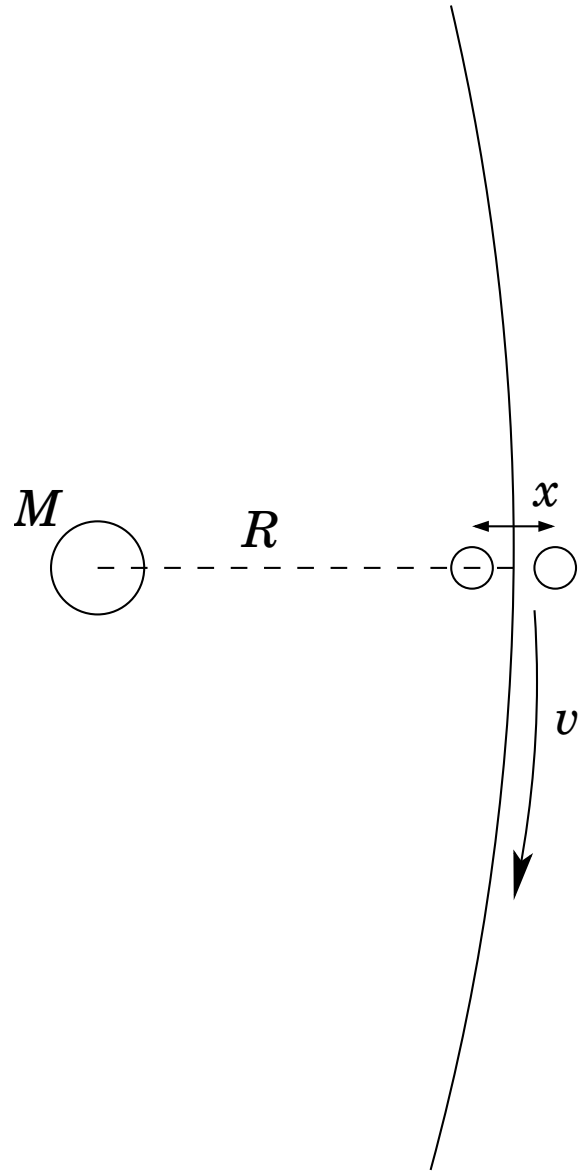}
\caption{Schematic of an encounter of two clouds in a cluster.  The clouds are
separated by a distance $x$.  The cloud complex model is located a distance $R$ from
the cluster core of mass $M$, moving at a velocity $v$ relative to the cluster.\label{clusterinteraction}}
\end{center}
\end{figure}

Estimates of the tidal effects can be made by considering 
two nearby clouds in the vicinity of a cluster
(Figure~\ref{clusterinteraction}). 
The two clouds will disperse at a rate 
$\Delta v=g\Delta t$ under the gravitational tidal acceleration $g$ 
from the cluster potential, where $g$ is
\begin{equation}
g=\frac{2GM_{cluster}x}{R_{cluster}^{3}}
\end{equation}
where $M_{cluster}$ is the mass of the cluster within radius R$_{cluster}$, 
and $x$ is the initial separation of the two clouds.  The timescale of the encounter can be taken as 
$\Delta t\sim R_{cluster}/v_{c,3d}$, where $v_{c,3d}=v_c/\cos \theta$ is the 
velocity of the cloud complex with respect to the cluster.  Taking 
$\Delta v=g\Delta t$, we can write a simple relation for the separation rate 
of the two clouds
\begin{equation}
\Delta v \approx\frac{2GM_{cluster}x}{R_{cluster}^2 v_{c,3d}}
\end{equation}
Two clouds initially separated by 10 kpc, located 845 kpc from the center of the 
cluster of mass $10^{14}$~\msun~and moving at a velocity of $600/\cos {\theta}$ 
\kms ~with respect to the cluster would roughly double their separation in a time
comparable with the cluster crossing time. Tidal forces related to the cluster
potential alone are unlikely to account for the spatial dispersion of the clouds 
in the complex, if they were stripped from a single galaxy.

The cloud complex is separated by $\Delta r\sim 235/\sin \theta$ kpc and by 
$\Delta v\sim 100/\cos \theta$ \kms ~in velocity from NGC 4424; the corresponding
numbers for NGC~4445 are respectively $\Delta r\sim 150/\sin \theta$ kpc 
$\Delta v\sim 200/\cos \theta$ \kms. This separation would increase from zero 
to the observed value over a time of order
\be 
\Delta t_{sep,4424}\sim 2.3(\tan{\theta})^{-1} \,\,\,\rm{Gyr}\label{cgsep4424}
\ee
for NGC~4424 and
\be 
\Delta t_{sep,4445}\sim 0.7(\tan{\theta})^{-1} \,\,\,\rm{Gyr}\label{cgsep4445}.
\ee
for NGC~4445. For relatively large values of $\theta$, they can be comparable 
with the time it takes for the galaxy to cross the inner regions of the cluster
$R_{cluster}/v_{gal,3d}$. Under constant acceleration, the time necessary to 
reach the observed values of $\Delta v$ and $\Delta r$ would be about twice 
$\Delta t_{sep}$ as computed above. Either of those assumptions are unrealistic 
if ram pressure acts upon the clouds, as they travel through an ICM of variable
density. However, two preliminary conclusions can be made already: (i) if the
clouds originated from NGC 4445 or NGC~4424, the stripping took place at least
a few $10^8$ yrs ago; (ii) simple kinematic timescales make NGC~4445 as a more
palatable candidate as a parent galaxy than NGC~4424. In the following, we 
concentrate our discussion on the assumption that the clouds were stripped from
NGC~4445, although we maintain the inferred relations parametrized form for easy application
to NGC~4424. Next, we derive the acceleration due to ram pressure.

As a spherical cloud of radius $r_c$ travels through the ICM at speed 
$v_{c,3d}=v_c/\cos \theta$, the force 
acting on its surface can be written as the ram pressure times the area, given by 
\be
F = n_{icm} m_{p} v_{c,3d}^{2} \pi r_{c}^{2}.
\ee

The resulting acceleration due to ram pressure is
\be
a_{rp} = \Delta v / \Delta t = \frac{\rho_{icm}v_{c,3d}^2 \pi r_{c}^2 \eta}{ M_{HI}}
\ee
where the factor $\eta$ accounts for the fraction of gas in form other than HI
and $M_{HI}$ is the HI mass of the cloud.
The acceleration can then be written as
\be
\label{eq:accel}
a_{rp} = 2.5 \times 10^{-9}\eta \left(\frac{n_{icm}}{10^{-3}~\rm{cm^{-3}}}\right) 
\left(\frac{r_{c}}{\rm{kpc}}\right)^{2} \left(\frac{v_{c,3d}}{10^3 ~\rm{\kms}}\right)^2
\left(\frac{M_{HI}}{10^8 ~M_{\odot}}\right)^{-1}~~\rm{[cm~s^{-2}]} 
\ee
Once the gas is stripped, while the motion of the galaxy is relatively unimpeded by the 
ram pressure, the clouds fall behind, affected by the acceleration $a$. In order to accumulate 
a velocity difference from the galaxy $\Delta v$, at constant acceleration
a time on order of $\Delta t_v = \Delta v/a$
will be required, while in order to accumulate a separation $\Delta r$, a
time on order of $\Delta t_r = (2 \Delta r/a)^{1/2}$ will be required:
\be
\Delta t_v = 1.3\times 10^8 ~\rm{yr}\,\left(\frac{\Delta v}{10^2 ~\rm{\kms}}\right)
A \cos \theta 
\label{dtv}
\ee
\be
\Delta t_r = 5.0\times 10^8 ~\rm{yr}~ \left(\frac{\Delta r}{10^2 ~\rm{kpc}}\right)^{1/2} 
A^{1/2}\cos \theta \sin^{-1/2} \theta
\label{dtr}
\ee
where
\be
A=\eta^{-1} 
\left(\frac{n_{icm}}{10^{-3}~\rm{cm^{-3}}}\right)^{-1}
\left(\frac{v_{c}}{10^3 ~\rm{\kms}}\right)^{-2}
\left(\frac{r_{c}}{\rm{kpc}}\right)^{-2} 
\left(\frac{M_{HI}}{10^8~M_{\odot}}\right)
\ee
For cloud C1, $M_{HI}=1.6\times 10^8$ \msun. The maximum radius out to which 
the VLA detects HI emission in C1 is about $1'.3$, which translates to 6.5 kpc at 
the Virgo cluster distance. However, the {\it effective}
radius of the cloud is very likely smaller than 6.5 kpc. A cursory inspection of 
Figure~\ref{C1moments}~shows that only about 1/3 of the area subtended by the emission out to a 1\arcmin~
radius exceeds a column density of gas at half the peak level, which would yield
an effective radius of order of 3 kpc. In addition, from 
the topology of the HI gas in our own ISM, it is found that while the majority
of the HI emission arises in denser regions, those clouds' volume filling factor
is quite small. Such structure would be unresolved by the 22\arcsec~(1.8 kpc) beam of
the VLA observations. Values of order $r_c=2$--4 kpc appear appropriate. The 
parameter $\eta$, which accounts for the mass fraction of He and other than HI gas, 
should be in the range 0.4--0.7. Near the current location of the clouds,
the ICM density is unlikely to exceed $n_{icm}\simeq 10^{-4}$ cm$^{-3}$, unless the 
ICM is very clumpy, as suggested by Kenney \etal ~(2004). As for the parameters
specific for the galaxies, for NGC 4424 $v_c=700$ \kms, $\Delta r=235$ kpc and
$\Delta v=100$ \kms, while for NGC~4445 $v_c=800$ \kms, $\Delta r=150$ kpc and
$\Delta v=200$ \kms. Note that by imposing 
$\Delta t_v \simeq \Delta t_r \simeq 2\Delta t_{sep}$ we can obtain a constraint 
on the combination of physical parameters 
\be
0.26 ~A^{1/2} \left(\frac{\Delta v}{10^2 ~\rm{\kms}}\right)
\left(\frac{\Delta r}{10^2 ~\rm{kpc}}\right)^{1/2} 
\sin^{1/2}\theta =1
\label{eq:equality}
\ee
As we discussed earlier, accumulation of distance and velocity offsets between clouds
and parent galaxy, $\Delta r$ and $\Delta v$, at either constant rate or constant 
acceleration are used under outlined above, and do not take
into account projection effects, which would account for the
orientation of NGC 4445 and the clouds with respect to M87 at the
cluster center.  It is important to note that a connection between NGC
4445
and the cloud complex
is somewhat ambiguous, especially when compared to other disrupted gas
features like NGC 4388 which show a clear connection in the form of a tail.


\subsection{Evaporation Timescale}

According to our findings, if the clouds were stripped from a parent 
galaxy, most of the stripping took place a fair fraction of a Gyr ago. In that
case, it is reasonable to investigate the survival of the clouds in the ICM.
Differential ram pressure forces will lead to ablation and dilution of the gas
clouds. In addition, upon removal from the galaxy, the cold gas can be heated 
by conduction by the intracluster gas and the gas mass of the cloud would thus
be progressively reduced by evaporation, if (presumably absent) stellar 
mass loss does not replenish the gas. The evaporation rate in \msun ~yr$^{-1}$ 
can be written as (Cowie \& Songaila 1977)
\be
\dot{M_{ev}}\simeq 16\pi \mu m_p \,\kappa \,r_{kpc}/25 k\simeq 35\, T_8^{5/2}\, r_{kpc} 
\,(40/\ln \Lambda)\,\,\, M_\odot\,\,\rm{ yr}^{-1}
\label{eq:mev}
\ee
where $\mu$ is the mean molecular weight, $m_p$ the proton mass, $\kappa$ the
thermal conductivity, $T_8$ the intracluster gas temperature in units of $10^8$ K,
$r_{kpc}$ the effective cloud radius in kpc (proportional to the square root of 
the area of the cloud exposed to the ICM), $k$ the Boltzmann constant and ln~$\Lambda$ the 
Coulomb logarithm. For $T_8 = 0.235$ (Shibata \etal ~2001) and $40/\ln \Lambda\sim 1$,
the evaporation timescale is
\be
\label{eq:tev}
t_{ev} = M_{gas}/\dot{M_{ev}} \geq 1.7\times 10^8 r_{kpc}^{-1} \eta^{-1}\psi^{-1}~~\rm{yr,}
\ee
where $\psi\leq 1$ is a correction factor roughly proportional to the fraction of 
the time the cloud's orbit -- between the time the gas is removed from the disk
and the time of observations --- places it in a part of the ICM within which the 
conductivity coefficient yields the evaporation rate in Eqn. \ref{eq:mev}.  
This estimate is very uncertain due to the largely unknown geometry of 
the cloud, an exact knowledge of its orbit through the ICM, the unknown fraction 
of the gas in ionized form and the possible 
presence of magnetic fields, which would inhibit conduction across field lines.
It is clear from Eqn. \ref{eq:tev} that if the HI gas is stripped from the galaxy
in a region deep enough in the cluster ICM, conduction can obliterate evidence
of the cold gas on timescales of $<10^8$ yr. However, the rate in Eqn. \ref{eq:tev} is 
approximately valid only in vicinity of the cluster core. Away from the central 
and denser parts of the cluster conduction becomes decreasingly effective, 
due to the rapid increase of the mean free path of electrons and conduction becoming 
saturated: this is thought to take place a couple of cluster ICM core radii 
($\sim 50$ kpc) out from the center of the ICM distribution (Sarazin 1986).
This is why in the simulation described in the preceding section, we have 
maintained the closest approach of the galaxy to M87, $r_p$, at values greater 
than a few hundred kpc; the evaporation timescale remains higher than the
kinematical ones and the clouds can be preserved. 


\section{Summary}
\label{summary}

We have utilized datasets from the ALFALFA survey and follow-up observations with 
the Very Large Array to obtain both single dish and aperture 
synthesis observations of a new HI cloud complex in the Virgo Cluster. 
The results of these observations are summarized as follows:

\begin{enumerate}
\item Five separate HI clouds in the ALFALFA survey have been discovered with
radial velocities between $cz_{\odot}~\sim$500 and 600 \kms. The complex subtends 
an angle of 35' (170 kpc), the individual cloud HI masses 
range  from 0.48 to 1.7 $\times 10^8$ \msun ~and the overall HI mass of the complex 
is $5.1\times 10^8$ \msun, at the distance of the Virgo cluster. The clouds'
velocity widths vary between 50 and 250 \kms. In the latter case, the wide
spectrum is likely to be due to a blend of several poorly resolved clumps.

\item Several cluster galaxies are found in the vicinity of the cloud complex;
the most likely candidates for association with the clouds are NGC~4424, a
disturbed SBa with $cz_{\odot} = 437$~\kms~at a projected distance of 235 kpc 
from the center of the cloud complex, and NGC~4445, an edge on spiral at a
projected distance of 150 kpc, with $cz_{\odot} = 354$~\kms. Both galaxies are
extremely HI deficient, apparently having lost between 80\% and 93\% of their
HI gas. The total HI mass of the cloud complex is equivalent to between 1/2
and 1/3 of the HI lost by NGC~4424 and about 4/5 of the HI lost by NGC~4445.

\item Two of the clouds in the complex (C1 ad C2) have been detected with the VLA. They
have angular sizes of 2.7\arcmin~and 1.5\arcmin~respectively. The synthesis maps
do not exhibit any degree of symmetry in the gas distribution or in the
velocity field. Faint optical features are found in the vicinity of the two
clouds. Lack of optical redshifts prevent us from establishing a physical
association. If the optical features were associated with the two clouds,
their HI mass to $g$--band luminosity ratios would be respectively $M_{HI}/L_g\sim 170$
for C1 and $M_{HI}/L_g\sim 30$ for C2, in solar units.
A handle on the HI size allows estimates of the dynamical masses of the clouds,
contained within the HI radius and under the hypothesis that they are self--gravitating.
Those values are $2.3\times 10^9$ \msun ~for C1 and $0.3\times 10^9$ \msun ~for C2.

\item The possibility that the clouds constitute a group of primordial structures,
embedded in their own dark matter halos, appears unlikely. A more plausible 
scenario is that the complex is material removed from a galaxy traveling through
the cluster at high speed, such as NGC~4424 or NGC~4445. A ram pressure stripping
event is preferred to a purely gravitational one.
    
\item A simulation of plausible orbital parameters for the putative parent galaxy
was carried out showing that a nearly radial orbit that would take it to within
400 kpc from M87 would produce effects comparable with those
observed. Nearest approach to M87 would have resulted  $\sim$ 1 Gyr ago, at which
stripping of most of the gas is assumed to have taken place. The nearest approach to the cluster center
would be approximately 500 kpc.  At that distance from 
M87, thermal conduction would be ineffective at 
evaporating the clouds after removal from the parent galaxy. At pericluster passage, 
the velocity (not corrected for l.o.s. inclination) of the galaxy would be 1100~
\kms. The inclination of the orbital plane to the line of sight of the best fit
simulation would be approximately 45$^{\circ}$. 

\item Dynamical timescales, based on relative velocities and spatial 
displacements of the clouds from the putative parent galaxy, suggest that
the most likely such parent is NGC~4445. 

\end{enumerate}

This research has made use of the NASA/IPAC Extragalactic Database (NED) which is 
operated by the Jet Propulsion Laboratory, California Institute of Technology, 
under contract with the National Aeronautics and Space Administration.   $Skyview$ was developed 
and maintained under NASA ADP Grant NAS5-32068
under the auspices of the High Energy Astrophysics Science Archive Research Center at the 
Goddard Space Flight Center Laboratory of NASA.  
KS acknowledges support from a Jansky Fellowship during the completion of this work.
This work has been supported by NSF 
grants AST--0307661, AST--0435697, AST--0607007 and the Brinson Foundation.

This research has made use of Sloan Digital Sky Survey (SDSS) data. Funding for the SDSS 
has been provided by the Alfred P. Sloan Foundation, the Participating Institutions, the National Aeronautics 
and Space Administration, the National Science Foundation, the U.S. Department of Energy, the Japanese Monbukagakusho, 
and the Max Planck Society. The SDSS Web site is http://www.sdss.org/.  The SDSS is managed by the Astrophysical Research 
Consortium (ARC) for the Participating Institutions. The Participating Institutions are The University of Chicago, Fermilab, 
the Institute for Advanced Study, the Japan Participation Group, The Johns Hopkins University, the Korean Scientist Group, Los
Alamos National Laboratory, the Max-Planck-Institute for Astronomy (MPIA), the Max-Planck-Institute for Astrophysics (MPA), 
New Mexico State University, University of Pittsburgh, University of Portsmouth, Princeton University, the United States 
Naval Observatory, and the University of Washington.






\newpage

\begin{deluxetable}{lc}
\tablecaption{ALFALFA Observing and Data Cube Parameters \label{AOobs}}
\tablewidth{0pt}
\tablehead{
	 \colhead{Parameter} & 
	 \colhead{Value}
}
\startdata
Sky center (J2000)                                &      $12\,20\,00$, $+09\,00\,00$\\
Spectral range	                          &      25 MHz (-2000 -- 3200 \kms) \\
Effective integration time                         &      48 seconds (beam solid angle)$^{-1}$\\
Spectral resolution \chan\                 &      24.4 kHz (5.1 \kms)\\
Half-power beam size                               &      $3\arcmin.3\ \times 3\arcmin.8$ \\
RMS noise $\sm$ for $\chan = 5.1\,$\kms           &      2.5 mJy/beam\\
\enddata
\end{deluxetable}

\begin{deluxetable}{ccccccc}
\tablecaption{Single-Dish Cloud Properties from ALFALFA Survey\label{AOparams}}
\tablewidth{0pt}
\tabletypesize{\footnotesize}
\tablehead{
            \colhead{Cloud} & 
	    \colhead{$\alpha,\delta$} & 
	    \colhead{\czsun} & 
	    \colhead{\wf} & 
	    \colhead{\fc} &
	    \colhead{\sn} &
	    \colhead{$\log(\mhi/\msun)$}
	    \\ 	    
	    \colhead{ } & 
	    \colhead{(J2000)} & 
	    \colhead{(\kms)} & 
	    \colhead{(\kms)} & 
	    \colhead{(Jy \kms)}  & 
	    \colhead{ } & 
	    \colhead{ } 
	    \\	    
	    \colhead{(1)} & 
	    \colhead{(2)} & 
	    \colhead{(3)} & 
	    \colhead{(4)} & 
	    \colhead{(5)} & 
	    \colhead{(6)} &
	    \colhead{(7)} 
}
\startdata
C1  & $12\,30\,25.8$, $+09\,28\,01$ &  $488 \pm 5$ & $62 \pm 11$ & $2.48 \pm 0.07$ & $21.2$ & $8.21$ \\
C2  & $12\,31\,19.0$, $+09\,27\,49$ &  $607 \pm 4$ & $56 \pm 7$  & $0.72 \pm 0.06$ & $6.5$  & $7.67$ \\
C3  & $12\,29\,42.8$, $+09\,41\,54$ &  $524 \pm 7$ & $116 \pm 15$& $1.16 \pm 0.07$ & $8.6$  & $7.87$ \\
C4  & $12\,30\,19.4$, $+09\,35\,18$ &  $603 \pm 4$ & $252 \pm 7$ & $2.56 \pm 0.09$ & $13.1$ & $8.22$ \\
C5  & $12\,31\,26.7$, $+09\,18\,52$ &  $480 \pm 10$& $53 \pm 21$ & $0.91 \pm 0.06$ & $7.6$  & $7.77$ \\ 
\enddata
\tablecomments{Col.~(1): cloud name. Col.~(2): right ascension and declination of cloud centroid (J2000). Col.~(3): average heliocentric velocity of integrated spectral profile from Figure~\ref{AOspectra}. Col.~(4): profile width, measured at 50\% of the integrated spectra profile peak and corrected for instrumental broadening as described in Giovanelli \etal (2007). Col.~(5): total flux of integrated spectral profile. Col.~(6): signal-to-noise ratio of the detection, computed using \wf\ and \fc\ via eq.~\ref{snr}. Col.~(7): base 10 logarithm of total HI mass, computed using \fc\ via eq.~\ref{HImass}.
}
\end{deluxetable}

\begin{deluxetable}{lc}
\tablecaption{Aperture Synthesis Observing and Data Cube Parameters \label{VLAobs}}
\tablewidth{0pt}
\tablehead{ 
	\colhead{Parameter} & 
	\colhead{Value}
}
\startdata
    Pointing center (J2000)                 & $12\,30\,45$, $+09\,26\,00$    \\
    Total time on-source                    &         265 min               \\
    Net bandpass                            &  2.2~MHz (378 -- 822~\kms)    \\
    Maximum spectral resolution \chanp      &    48.8 kHz (10.3~\kms)       \\
    Natural Weighting:                      & \\
       \hspace{8pt} Synthesized beam        & $22.2'' \times 20.7''$ @ \,-20\dg\\
       \hspace {8pt} $\smp$ at pointing center, $\chanp =20.7\,$\kms\      &      0.33 mJy/beam\\
       \hspace {8pt} $\smp$ at pointing center, $\chanp =31.0\,$\kms\      &      0.29 mJy/beam\\
\enddata
\end{deluxetable}

\newpage


\begin{deluxetable}{cccccccccc}
\tablecaption{Aperture Synthesis Cloud Properties from the VLA Observations \label{VLAparams}}
\tablewidth{0pt}
\tabletypesize{\footnotesize}
\rotate
\tablehead{ 
	\colhead{Feature} & 
	\colhead{$(\alpha,\delta)^\prime$} & 
	\colhead{\czsunp } & 
	\colhead{\wfp } & 
	\colhead{\fcp } & 
	\colhead{\ahip} &
	\colhead{\pahip} &
	\colhead{$\log(\mhip/\msun)$} & 
	\colhead{$\log(\mdynp/\msun)$}  
	\\
	\colhead{} &
   \colhead{(J2000)} & 
   \colhead{(\kms)} & 
    \colhead{(\kms)} &  
    \colhead{(\jykms)} & 
    \colhead{(\arcmin)} &
    \colhead{(\arcdeg)} &
    \colhead{} & 
    \colhead{}
    \\
    \colhead{(1)} & 
    \colhead{(2)} &  
    \colhead{(3)} & 
    \colhead{(4)} &  
    \colhead{(5)} & 
    \colhead{(6)} & 
    \colhead{(7)} &
    \colhead{(8)} &
    \colhead{(9)}
}
\startdata
 C1 &  $12\,30\,24$ $+09\,28\,20$ & $488 \pm 6$ & $78 \pm 11$ & $2.14 \pm 0.07$ & $2.7 \pm 0.3$  & 36 & $8.15$ & $9.36$ \\
 C2 &  $12\,31\,18$ $+09\,29\,25$ & $597 \pm 3$ & $38 \pm 5$ & $0.57 \pm 0.06$  & $1.5 \pm 0.3$  & 146 & $7.57$ & $8.48$ \\
\enddata
\tablecomments{Col.~(1): cloud name. Col.~(2): right ascension and declination of the peak \nhp\ (J2000). Col.~(3): average heliocentric velocity of integrated spectral profile from Figure~\ref{VLAspectra}. Col.~(4): profile width, measured at 50\% of the integrated spectral profile peak and corrected for instrumental effects assuming that unbroadened profile is gaussian. Col.~(5): total flux of integrated spectral profile. Col.~(6): maximum linear extent of region with $\nhp \geq 10^{20}~\cm$ in the total intensity maps (Figs~\ref{C1moments}a~and~\ref{C2moments}a). Col.~(7): position angle at which \ahip\ was measured. Col.~(8): base 10 logarithm of total \hi\ mass, computed using \fcp\ via eq.~\ref{HImass}. Col.~(9): base 10 logarithm of the dynamical mass, computed using \wfp\ and \ahip\ via eq.~\ref{Dynmass}.
}
\end{deluxetable}


\clearpage

\parskip 8pt

\clearpage

\clearpage

\end{document}